\documentclass[12pt]{article}

\usepackage{amsmath,amsfonts}
\makeatletter \@addtoreset{equation}{section}

\usepackage{cite}
\usepackage{bm}
\usepackage{dcolumn}
\usepackage[pdftex]{graphicx}
\usepackage{wrapfig}
\usepackage{sidecap}
\usepackage{graphicx}
\usepackage{caption}
\usepackage{subcaption}
\usepackage{graphicx}
\usepackage{dcolumn}
\usepackage{eufrak}
\usepackage{yfonts}
\usepackage{dsfont}
\usepackage{bbold}
\DeclareMathAlphabet{\mathpzc}{OT1}{pzc}{m}{it}
\makeatother
\def\one{{\hbox{ 1\kern-.8mm l}}}
\newcommand{\Dslash}{\not{\hbox{\kern-4pt $D$}}}
\newcommand{\pdslash}{\not{\hbox{\kern-2pt $\partial$}}}
\newcommand{\be}{\begin{equation}}
\newcommand{\bea}{\begin{eqnarray}}
\newcommand{\eea}{\end{eqnarray}}
\newcommand{\ba}{\begin{array}}
\newcommand{\ea}{\end{array}}
\newcommand{\ee}{\end{equation}}

\begin{document}

\begin{titlepage}
\vspace*{1mm}%
\hfill%

\vspace*{15mm}%
\begin{center}

{{\Large {\bf Violation of Invariance of Measurement for GDP Growth Rate and Its Consequences}}}

\vspace*{15mm} \vspace*{1mm} {Ali Hosseiny  }

 \vspace*{1cm}

{\it  Department of Physics, Shahid Beheshti University \\
G.C., Evin, Tehran 19839, Iran. }\\
email: \tt al$\_$hosseiny@sbu.ac.ir

\vspace*{2cm}

\end{center}

\begin{abstract}

The aim here is to address the origins of sustainability for the real growth rate in the United States. For over a century of observations on the real GDP per capita of the United States a sustainable two percent growth rate has been observed. To find an explanation for this observation I consider the impact of utility preferences and the effect of mobility of labor \& capital on every provided measurement. Mobility of labor results in heterogenous rates of increase in prices which is called Baumol's cost disease phenomenon. Heterogeneous rates of inflation then make it impossible to define an invariant measure for the real growth rate. Paradoxical and ambiguous results already have been observed when different measurements provided by the World Bank have been compared with the ones from the systems of national accounts (SNA). Such ambiguity is currently being discussed in economy. I define a toy model for caring out measurements in order to state that this ambiguity can be very significant. I provide examples in which GDP expands 5 folds while measurements percept an expansion around 2 folds. Violation of invariance of the measurements leads to state that it is hard to compare the growth rate of GDP for a smooth growing country such as the U.S. with a fast growing country such as China. Besides, I state that to extrapolate the time that economy of China passes the economy of the US we need to consider local metric of the central banks of both countries. Finally I conclude that it is our method of measurements that leads us to percept the sustainable growth rate.

\end{abstract}

\end{titlepage}

 \tableofcontents

 %--------------------------------------------------------------------
\section{Introduction}

In an influential paper in 1957, Nicholas Kaldor counted six historical facts which he notified that growth models should be able to address:\\

1. The rate of return to capital has been almost constant over the time

2. Labor share of income is sustainable in the long run

3. The ratio of capital to output has kept a sustainable rate in the long run

4. Capital per labor has grown with a constant rate

5. Real wages has sustained an almost constant growth rate over the time

6. Productivity of labor has grown with  an almost constant rate in long periods of time\\

One of these facts is sustainable growth rate or the sixth fact in our list. The sixth fact indicates that  although the annual growth of GDP per capita has serious fluctuations, in the long run it is about 2\%.
 While exogenous growth models have been successful to address Kaldor's five other facts, they have not been able to provide explanation concerning the sixth fact. 
 
Before proceeding it is worth noting that productivity is a consequence of technology improvements. Now doesn't an effective constant two percent growth of technology for over a century seem strange? 
 In this paper I aim to present an explanation for this strange observation. In this regard I first emphasize that measurement of the growth rate is not invariant and strongly depends on the pattern of growth. In other words, I notify that measurements of growth for two countries with exactly initial and final conditions might give quite different results. Dependence of the measurement on the pattern of growth is not an unknown phenomenon. In this paper however through studying toy models and running hypothetical experiments I carry out  an examination on the size of inconsistency in measurements. After that I go through mathematical equations and find the value of sustainable growth rate in terms of other macroeconomical variables. In other words, I provide equations to examine the influence of the other five facts on this fact.

The invariance problem concerning measurements of GDP has recently attracted a lot of attention. Increasing attention in the recent years has been due to the observation of paradoxical results concerning international data. When Penn World Table or World Bank's reports\footnote{International Comparison Program or ICP} on GDP of different countries were compared to the ones from the system of national accounts, some big inconsistencies were observed. The Penn World Table or World Bank's report provides basis for finding the GDP of different countries through Purchasing Power Parity (PPP) basis. When we aim to compare GDP of different countries through PPP it means that we measure GDP of all countries through a standard price base.

 Suppose that we have compared GDP of two countries a couple of years ago through the World Bank report. To compare their current GDP we have two choices; one choice is to consider their old GDP, and via the data on national accounts and growth rates, extrapolate their current GDP, the other choice is to look at an updated report of the World Bank and compare their current GDP size. It has been observed that surprisingly these two methods lead to paradoxical results. This inconsistency has attracted a lot of attention, see e.g. Crawford et al. 2008, Deaton et al. 2010, 2014, Feenstra et al. 2013a, Prasada et al. In some papers the problem has been called "space-time inconsistency" while space means measurement in different countries, See for example Feenstra 2009 \& 2013b and Oulton 2015. In this paper we look at the space time inconsistency for measurement of GDP growth rate and some of its probable consequences.

 Gross domestic product by definition is the aggregate measure of all final goods and services produced in a state in a period of one year. In other words we can define GDP in the year $i$ as
\bea
GDP_i=\Sigma_aY^i_aP^i_a,
\eea
in which $a$ stands for any form of final good or service provided in a state, $Y^i_a$ stands for its quantity, and $P^i_a$ stands for its price in that year. Prices themselves are subject to the equilibrium conditions and may grow with different rates. Now, for measuring and comparing the GDP of different years, or comparing the GDP of different countries, a problem is faced. The problem occurs when aiming to aggregate the outputs ($Y^i_a$) for all of the sectors in the country. The reason lies in the fact that product prices are our meter for making this aggregation. This meter is not only non-stationary but also its dynamics for different sectors is heterogeneous. Now, there is a question to be answered: is our measurement for growth invariant in the space of productions? 

Actually, based on Baumol's cost disease phenomenon (Baumol and Bowen 1966), since different sectors in economy may face different levels of advancement in technology, as a result they can face different rates of inflation. The fact will be clearer when we compare two sectors from service side and manufacturing side. Since usually sectors in service side have lower rates of improvement, they face higher rates of growth in prices. Many observations and reports have been provided in Baumol (2012). For example we observe that based on the report issued by the National Center for Education Statistics 2010 while in 1980 it cost \$3500 per year in average to attend a bachelor program in the US, in 2008 the rate was about \$20,500. Such a report reveals a rate of inflation around 6\% for the education sector. This is while the average of inflation has been far below this rate. According to the National Center for Public Policy and Higher Education report issued in 2008, from the early 1980s, beside education which had faced a growth around \%440 percent during the studied period, expenses  for medical care had faced a growth as of \%250.\footnote{A higher rate of inflation for many other sectors in the service side has been observed. According to Baumol 2010 and based on the reports issued by the U.S. Bureau of Labor Statistics 2009a and 2009b such a phenomenon has been observed for lawyers' fees or even for funeral services.}\footnote{Significant increases in health expenses has been observed even in emerging economies as BRIC nations (Brazil, Russia, India, and China) according to Lu et al. 2010.} This was while the average growth was around \%110. If we think about the big portion of service sector in GDP we find that in order to have an average growth as of \%110, the manufacturing side should have a rate much below it. Therefore, existence of heterogeneous rates of inflation is clear and is being widely studied especially in fields which are concerned about services.
                 
Studying the effect of heterogeneous prices on measurements is not new and has been widely discussed mainly in the context of price index. When measuring inflation, beside heterogeneity problem we face substitution effect. As price of one good soars then agents may substitute a different good to consume. In this case, when we want to measure the cost of living, each year has its own basket of goods consumed. The substitution effect and the influence of different definitions for baskets of goods has been discussed widely in price index problem. Invariance of such measurements was first discussed by Irving Fisher in 1927. 
Later some important movements were addressed by Laspeyres, Paasche, Törnqvist or Fisher. For recent studies see Balk 1995, Diewert 1976, Feenstra et al. 2000 and Oulton 2008.

In this paper I look for the origins of sustainable growth rate. I conclude that perception of sustainable growth rate is a matter of measurement rather than a real fact. To explain, I review the production function and impose equilibrium conditions and relate dynamics of long run prices to technology. I then make a toy model to study the impact of heterogeneous advances in technology on structural changes and measurement for the growth rate. We observe that despite early visions, the growth rate could be perceived with quite different rates. I provide toy models of some bi-sector economies. In my toy models I provide an example of three different countries with exactly similar initial and final conditions. After 100 years, production in each sector grows 19  folds in all countries. This is while each country has a different pattern of growth. Though production finally grows 19 folds, based on annual measurements of real growth, a country percepts a growth as of 31 folds for 100 years while another country percept a growth equal to 8 folds. No wonder we observe paradoxical results when comparing GDP of different countries in real world. GDP is not a scalar and different meters suggest different distances between points. Once we observed that GDP of China shrunk substantially (See Feenstra et al. 2013-a). I then provide justification to conclude that perception of sustainable growth rate is mainly a result of violation of invariance of measurement. 

In my work beside heterogeneity in prices I consider structural change as a result of improvement in technology. The point is that as long as technology grows in some sectors, production relies more on capital rather than labor. As a result we have reallocation of labors and structural change. Growth rate, structural changes, and occurrence of Kaldor's facts as a result of cost disease have been of interest in studies such as Ngai et al. 2007, Foellmi et al. 2008, Herrendorf et al. 2013, and Boppart 2014. Though these models aim to relate dynamical prices and utility preferences to the sustainable growth rate, but their result mainly depends on properties of utility preferences. Similar to Foellmi and Zweimüller 2008, Engel's curves are important in our discusstions. I however have no hypothesis concerning R\&D. Improvment in technology can happen in all sectors. My work mainly relies on the matter of measurement. I do not impose narrow and special binds to utility preferences. As a result of my description, I claim that if labor income looses its historical share, the growth rate will accordingly change. Labor share is declining and my description suggests different growth rates in the future. 

This paper is organized as follows. In section II I review production function. I then review constraints on production function and their consequences on the long run prices and allocation of labors. In section III I define a toy model in which we can simulate and evaluate breakdown of invariance in the measurements of our toy model. In my toy models I provide examples in which two countries can have exactly initial and final conditions while during a period their central banks percept different growth rates. In section IV I go for further analysis and conclusion.

%%%%%%%%%%%%%%%%%%%%%%%%%%%%%%%%%%%%%%%%%%%%%%%%%%%%%%%%%%%%%%%%%%%%%%
%%%%%%%%%%%%%%%                                                                                     %%%%%%%%%%%%%%%
%%%%%%%%%%%%%%%       Section II{ Production Function
%%%%%%%%%%%%%%%                                                                                     %%%%%%%%%%%%%%%
%%%%%%%%%%%%%%%%%%%%%%%%%%%%%%%%%%%%%%%%%%%%%%%%%%%%%%%%%%%%%%%%%%%%%%

\section{Production Function, a Short Review}
My aim in this section is to relate long run prices to the level of technology quantitatively. I first review production function. I then consider a heterogeneous world and through writing production function in terms of physical capital try to address equilibrium price for goods and capital. Besides I review constraints that given utility functions are able to address allocation of labors and the size of sectors. I then consider a simple economy and find long run prices of goods and capital. This way in my simulation I can track long run prices, allocation of labors and the size of GDP at equilibrium.

%%%%%%%%%%%%%%%%%%%%%%%%%%%%%%%%%%%%%%%%%%%%%%%%%%%%%%%%%%%%%%%%%%%%%%
%%%%%%%%%%%%%%%                                                                                     %%%%%%%%%%%%%%%
%%%%%%%%%%%%%%%       Section II{ Production Function
%%%%%%%%%%%%%%%                                                                                     %%%%%%%%%%%%%%%
%%%%%%%%%%%%%%%%%%%%%%%%%%%%%%%%%%%%%%%%%%%%%%%%%%%%%%%%%%%%%%%%%%%%%%
\subsection{Production Function and Dynamics of Long Run Prices in a Heterogeneous World}

Let's at first review shortly the production function. It is a function which is supposed to indicate the quantity of output of a firm subject to be given the number of labors employed and capital invested in it.  The function is usually written as follows
\bea
Q_a=Y_a(T_aL_a,K_a),
\eea
in which $Q_a$ is the quantity of good $a$ produced, $L_a$ is the number of labors hired in the firm and $T_a$ indicates productivity level and grows as long that new technology is achieved.  Capital is indicated by $K_a$. The function is supposed to have scaling features as
\bea\label{scalingtwo}
Y_a(T_azL_a,zK_a)=zY_a(T_aL_a,K_a),
\eea
which results in
\bea\label{scaling}
Y_a(T_aL_a,K_a)=T_aL_aY_a(1,\frac{K_a}{T_aL_a})=T_aL_ay_a(1,k_a),
\eea
in which $k_a=\frac{K_a}{T_aL_a}$ is capital per effective labor. Beside scaling features, the production function is supposed to have diminishing return to capital. In other words it is supposed to be a convex up function respect to capital.  

In a competetive market in the long run we expect
\bea\begin{split}\label{wageproductionone}
&P_a\frac{\partial Y_a}{\partial L_a}=W,
\cr&P_a\frac{\partial Y_a}{\partial K_a}=R_c+\delta_a,
\end{split}\eea
in which $P_a$ indicates price for units of good $a$, $W$ stands for wage, $r$ is real interest rate and $\delta_a$ stands for depreciation of capital in sector $a$. 

 General understanding concerning Eq. (\ref{wageproductionone}) was that it indicated level of wage in each sector given the price set of products. Baumol and Bowen however in 1966 notified that firms are price taker for wage. Wages and rate of return on capital are the same for all sectors and it is long run price that should be adjusted so that Eq. (\ref{wageproductionone}) hold. 
Based on Baumol's cost disease phenomenon and as a result of equation (\ref{wageproductionone}) sectors that have higher rates of growth in $T_a$ face lower rates of inflation.

%%%%%%%%%%%%%%%%%%%%%%%%%%%%%%%%%%%%%%%%%%%%%%%%%%%%%%%%%%%%%%%%%%%%%%
%%%%%%%%%%%%%%%                                                                                     %%%%%%%%%%%%%%%
%%%%%%%%%%%%%%%       Section II{ Production Function
%%%%%%%%%%%%%%%                                                                                     %%%%%%%%%%%%%%%
%%%%%%%%%%%%%%%%%%%%%%%%%%%%%%%%%%%%%%%%%%%%%%%%%%%%%%%%%%%%%%%%%%%%%%
\subsection{Soup of Firms and Equilibrium Conditions in a More Complicated Economy}\label{mmm}
 
 For a more complicated world we need to write production functions in a more realistic way 
\bea\label{pfunctionanomaly}
Q_a=Y_a(T_aL_a,N_{ab}),
\eea
in which $N_{ab}$ is the number of units of good $b$ which is used as physical capital in sector $a$. Basically $b$ can stand for any form of good produced in the market. Money value capital in sector $a$ is $K_a=\Sigma_bP_bN_{ab}$.
\bea
P_a\Delta Y_a \ge P_b\Delta N_{ab}(R_c+\delta_b).
\eea
Hiring more labors is reasonable if and only if
\bea
P_a\Delta Y_a\ge W\Delta L_a.
\eea
After market clearing for all goods and variables including market for loanable funds and as well labor force inequalities in the above equations tend to equality.
\bea\label{capitalab}
\frac{P_a\partial Y_a} {P_b\partial N_{ab}}=(R_c+\delta_b),
\eea
and
\bea\label{laborab}
\frac{P_a\partial Y_a} {\partial L_a}=W.
\eea
These two equations are very critical and address long run prices which are our meter to measure growth rate. Let's focus on them more deeply. In an economy which have $M$ sectors which produce $M$ forms of intermediate and final goods and services, to address aggregate capital and aggregate production, the theory needs to identify the set of variables $N_{ab}$ which are $M^2$ variables. Beside, the theory needs to address long run price for all goods which are $M$ variables and as well distribution of labors. If we have set of variables $\{N_{ab},L_a,P_a\}$ known then we can find aggregate production and GDP. Equations (\ref{capitalab}) and (\ref{laborab}) set $M^2+M$ constraints on the system. They can address long run prices and as well $N_{ab}/L_a$ or capital per labor in each sector. Maximizing utility preferences provide $M$ further equations which can identify distribution of labors and GDP at equilibrium can be addressed subject to rate of return on capital. For a more detailed review and and as well discussion from a thermodynamics point of view see Hosseiny and Gallegati (2016).

%%%%%%%%%%%%%%%%%%%%%%%%%%%%%%%%%%%%%%%%%%%%%%%%%%%%%%%%%%%%%%%%%%%%%%
%%%%%%%%%%%%%%%                                                                                     %%%%%%%%%%%%%%%
%%%%%%%%%%%%%%%             \section{A misperception of GDP growth rate}
%%%%%%%%%%%%%%%                                                                                     %%%%%%%%%%%%%%%
%%%%%%%%%%%%%%%%%%%%%%%%%%%%%%%%%%%%%%%%%%%%%%%%%%%%%%%%%%%%%%%%%%%%%%

\section{A misperception of GDP growth rate}

Neoclassical growth models widely use the concept of aggregate production function: $Y_t(K_t,TL_t)$. Exogenous growth models take $T$ as an exogenous variable and supposes that it grows with a rate $g$. They have no explanation why $g$ should fluctuate around a sustainable rate. In other words it seems that by some strange reason, modern technologies are achieved in a manner that the aggregate production sustains its 2 percent growth rate. I believe that this effect is an outcome of our measurements. Let's look at measurement for GDP growth rate. By definition we can find the growth rate as
\bea\label{growthintro}
g_i=\frac{GDP_i-GDP_{i-1}}{GDP_{i-1}},
\eea
 in which $GDP_i$ is the gross product in year $i$, obtained by
\bea
GDP_i=\Sigma_aY^i_aP^i_a.
\eea
Note that $GDP_{i-1}$ is the gross product for the year prior to year $i$. We now face a problem. Price is a nominal variable which could increase over a two year period due to the central bank monetary policy. Therefore we may over estimate the growth rate. To solve this problem prices of a selected year is chosen as the base price or standard price for a period and the size of GDP all years within that period are measured with the same price base. Therefore the growth rate would be
 \bea\label{GDPgrowthintro}
g_i=\frac{GDP_i^j-GDP^j_{i-1}}{GDP^j_{i-1}}
\eea
in which $GDP^j_i$ is GDP in the $i_{th}$ year with the reference prices chosen from the $j_{th}$ year which is the base year. In other words $GDP^j_i$ is defined as
\bea
GDP^j_i=\Sigma_aY^i_aP^j_a,
\eea
where $P^j_a$ is the prices of product $a$ in the base year. Now we are left with a question. How seriously our measurements might be influenced by the price set. If we a had a uniform rate of inflation for all sectors, then we would have no problem. The bad news is that since the rates of inflation are heterogeneous, overcoming the problem is non-trivial. As a matter of fact there are a few expectations when dealing with the $real$ growth rates. These expectations are listed in as follows

1. If during the period of $\alpha$ years through technological advances, productivity for major sectors are multiplied by a factor $S$, then for the real growth average during this period we expect
\bea\label{property1}
\bar g=\frac{ln(S)}{\alpha}.
\eea
 As an example, if productivity of almost all sectors doubles in 10 years, we expect the growth rate to be $7\%$ in average.\\

2. If we want to compare real GDP of two years say $i_{th}$ and $j_{th}$ year we expect
\bea\label{property2}
GDP_i=[{\prod_{l=j+1}^i(1+ g_l)}]GDP_j,
\eea
in which $g_l$ is the real growth rate in the $l_{th}$ year, where both GDPs are measured with the same price base. \\

3. We expect our measurement for the real growth rate to be invariant under the pattern of growth. Let's consider two nations which in 1800 and 1900 both had exactly the same economy. I mean the same labor force, the same capital, the same amount of production in each sector of economy and the same level of technology for producing any form of goods or services. Now, suppose that though their initial and final conditions are the same but in the intermediate years they experience different patterns of growth. Since initial and final situations are the same we expect
\bea
\Sigma_{i=1800}^{1900} g_i=\Sigma_{i=1800}^{1900}g_i^\prime,
\eea
in which $g_i$ is the real growth in the $i_{th}$ year in one of the nations and $g_i^\prime$ is the same parameter for the other.\\

Since we have not defined an invariant method to measure the growth rate, then it is expected that all of these properties are violated. The first question is that to what extent could it happen? The second question is how equilibrium conditions through violating these equations leads us to percept a sustainable measure for the growth rate? In this section I consider some simple economies as toy models and run different patterns of growth and evaluate the strength of violation in my toy models.

To have a simple toy model I first suppose that we have two closed economies in two separate islands which we call North and South Island. I as well suppose that in each island the production is bread together with a luxurious service. I denote the first sector by $A$ and the second sector by $B$. I as well suppose that the production function for both sectors are Cobb-Douglas function
\bea\label{yyy}
Y_a=T_a^{\lambda_a}L_a^{\lambda_a}N_a^{1-\lambda_a},
\eea
in which $a$ stands for both sectors $A$ and $B$. I then suppose that the utility preference of the people of each island is
\bea\label{utilitymiperception}
U=(\frac{Y_A}{L_t}-N^0)(\frac{Y_B}{L_t})^{\Omega},
\eea
where we have $\Omega\ge1$. What this form of utility means? Based on this utility function a minimum of bread per capita is crucial for happiness. Otherwise utility is negative. When the minimum food has been fulfilled, then for extra outputs the luxurious service $B$ is more desired. If we set for example $\Omega=4$, then for economy which has a capacity to produce bread much more than $N^0$, then the utility will reduce to $U\approx(\frac{Y_A}{L_t})(\frac{Y_B}{L_t})^{4}$. So, in such situations a big portion of efforts goes to the sector $B$. Let's make this point more clearer through maximization of utility. I keep rate of return equal to $\%6$ and suppose that capital in each sector is maintained in a level that through Eq. (\ref{capitalab}) this profit is fulfill. Now, through Lagrange method we find that
\bea\begin{split}
&\frac{\partial U}{\partial L_A}=\frac{\partial U}{\partial L_B}\;\;\;\Rightarrow\;\;\;\frac{\partial U}{\partial Y_A}\frac{\partial Y_A}{\partial L_A}=\frac{\partial U}{\partial Y_B}\frac{\partial Y_B}{\partial L_B}\cr& \cr&\Rightarrow\lambda_AY_AL_B=\Omega\lambda_B(Y_A-L_tN^0)L_A.
\end{split}\eea
Plugging $Y_A$ from Eq. (\ref{yyy}) we find
\bea\label{labormispersepction}
L_A=L_t\frac{\lambda_A+\Omega\lambda_BN^0/T_A(\frac{1-\lambda_A}{R_c+\delta_A})^{\frac{1-\lambda_A}{\lambda_A}}}{\lambda_A+\Omega\lambda_B},\;\;\;\;\;\;\;\;\;\;\;L_B=L_t-L_A.
\eea

%%%%%%%%%%%%%%%%%%%%%%%%%%%%%%%%%%%%%%%%%%%%%%%%%%%%%%%%%%%%%%%%%%%%%%
%%%%%%%%%%%%%%%                                                                                     %%%%%%%%%%%%%%%
%%%%%%%%%%%%%%%             \section{A misperception of GDP growth rate}
%%%%%%%%%%%%%%%                                                                                     %%%%%%%%%%%%%%%
%%%%%%%%%%%%%%%%%%%%%%%%%%%%%%%%%%%%%%%%%%%%%%%%%%%%%%%%%%%%%%%%%%%%%%

\subsection{A Model to Violate Properties 1 and 2}

Back to our islands I suppose that each island has $100,000$ labor. For both islands in 1900 I suppose $T_A=T_B=1$ and I set depreciation rate for both sectors $\delta=5.5\%$. For the production function I suppose $\lambda_A=\lambda_B=2/3$. The only difference is utility preferences. In North Island I have $\Omega=1$ and in South Island I have $\Omega=5$. The utility preference in Eq. (\ref{utilitymiperception}) for North Island I suppose $N^0=1.6990$ and for South Island I suppose $N^0=1.6711$. Given these rates, for both countries $98\%$ of labors work in sector A. So, in both countries almost all of the attention is spent on sector $A$. Now, since $98\%$ of economy belongs to sector $A$ then it seems that if we let the productivity to grow annually with the rate of $5\%$ for sector A, then a real growth rate for GDP close to $5\%$ is expected. In other words since in a period of 98 years the productivity in sector $A$ which compensates for $98\%$ of the economy grows as much  as 119 folds, therefore it is expected that the real GDP would grow 119 folds in 98 years. 

\begin{figure}
        \centering
        \begin{subfigure}[b]{0.5\textwidth}
                \centering
                \includegraphics[width=\textwidth]{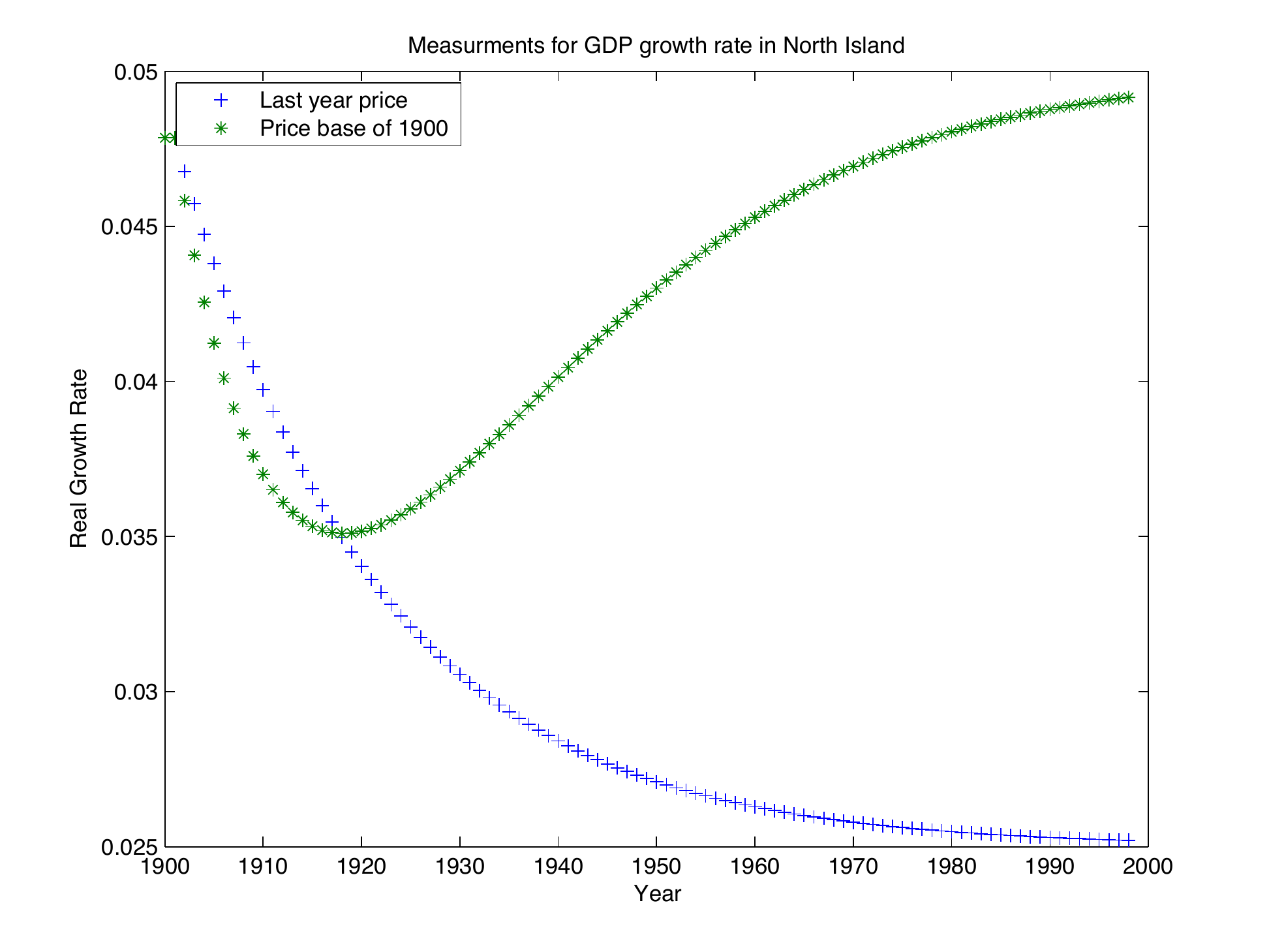}
                \caption{}
                \label{fig:gkull}
        \end{subfigure}%
        \begin{subfigure}[b]{0.5\textwidth}
                \centering
                \includegraphics[width=\textwidth]{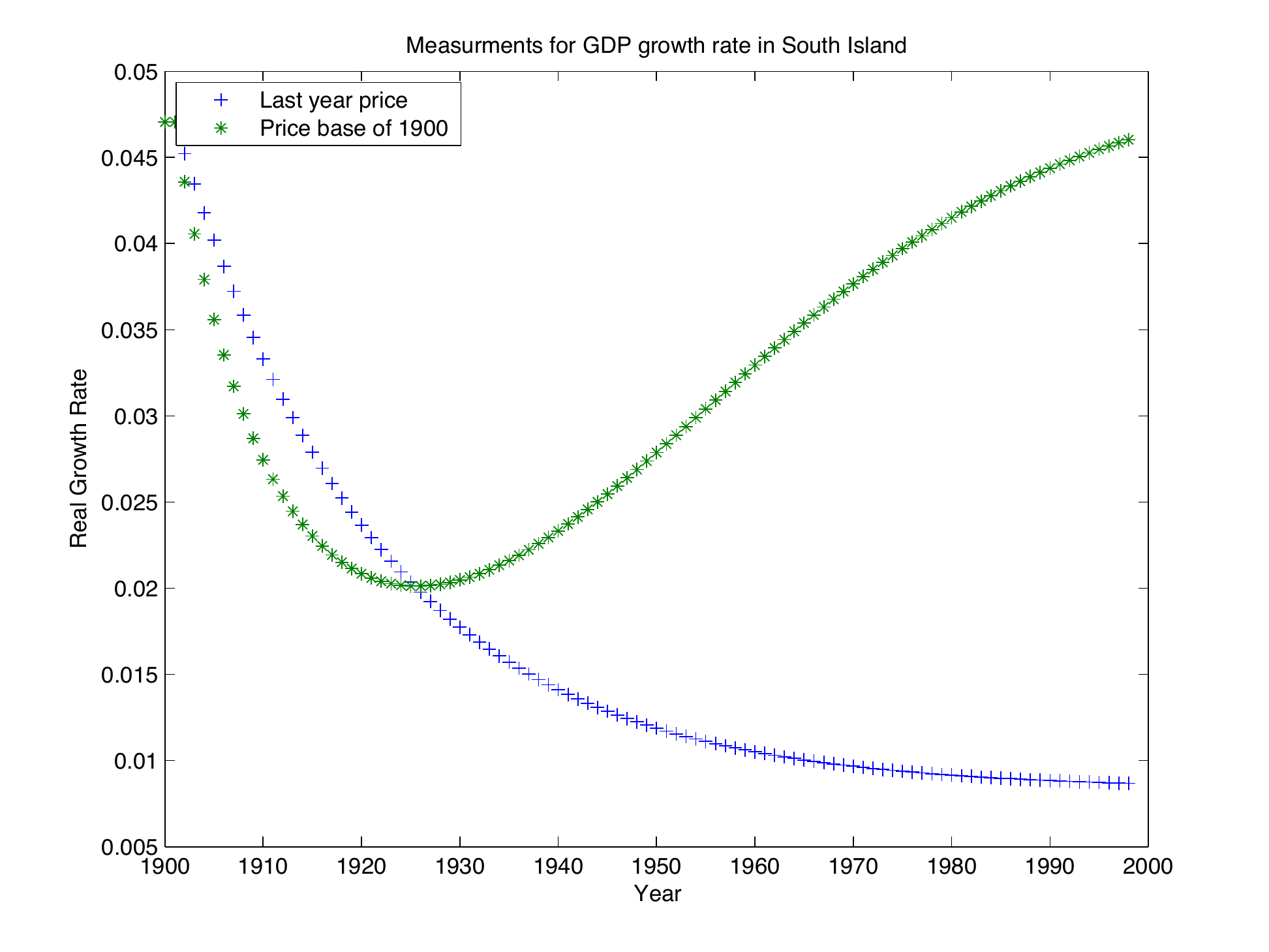}
                \caption{}
                \label{fig:tiger}
        \end{subfigure}
        
        \begin{subfigure}[b]{0.5\textwidth}
                \centering
                \includegraphics[width=\textwidth]{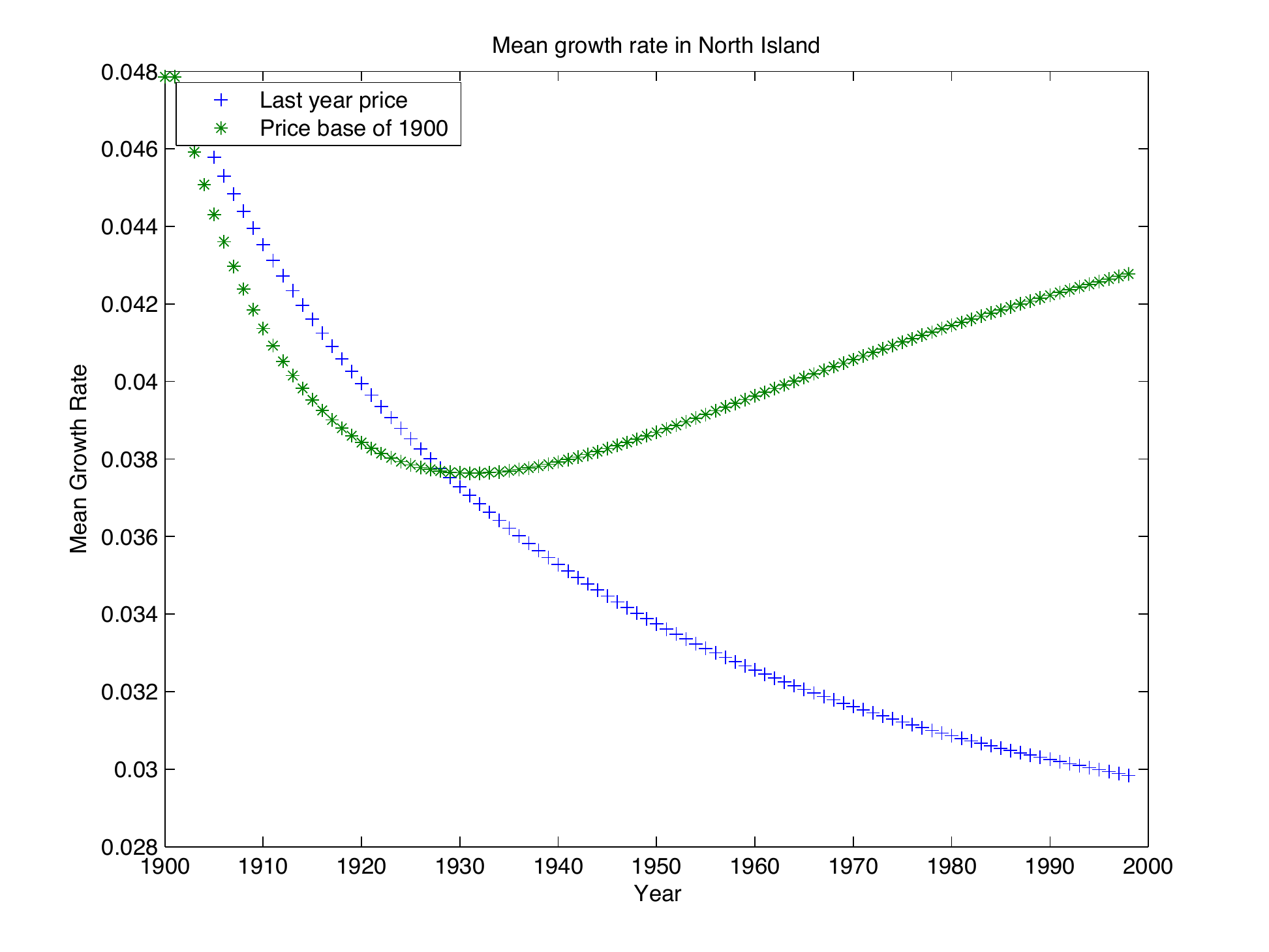}
                \caption{}
                \label{fig:kgull}
        \end{subfigure}%
\begin{subfigure}[b]{0.5\textwidth}
                \centering
                \includegraphics[width=\textwidth]{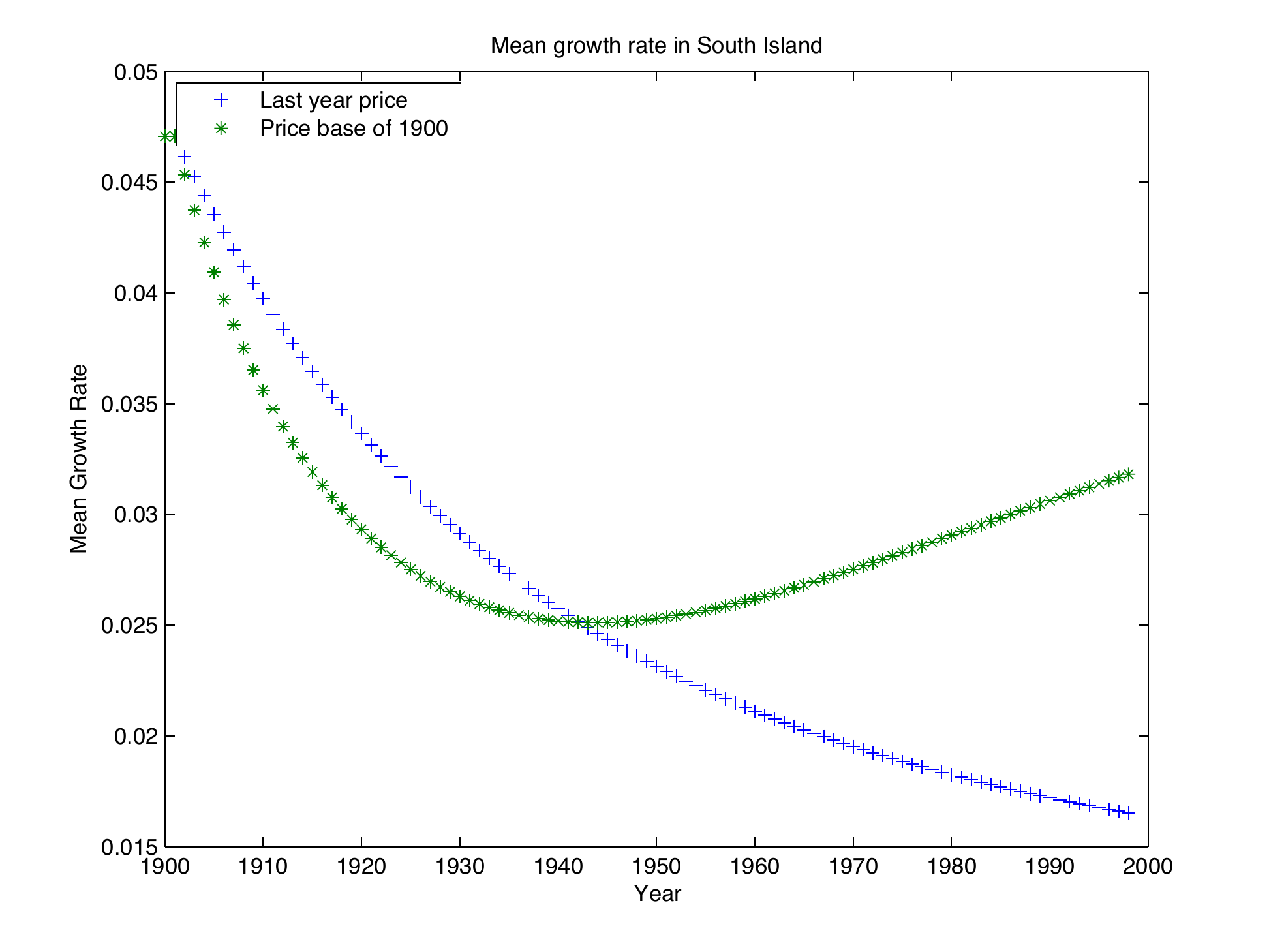}
                \caption{}
                \label{fig:tikger}
        \end{subfigure}

        \caption[font=footnotesize]{ Real growth rate for the North and South Islands in experiment I. The '+' sign represents measurements in which for each year, the previous year has been considered as the base year. The '*' sign shows measurements in which the year 1900 has been considered as the base year for the prices.
a) The real growth rate in the North Island, b) The real growth rate in the South Island, and c,d) The mean value of the growth rate from year 1900 to the desired year. As it can be seen, all results are seriously dependent to the base year. Besides, despite the 5 \% growth rate of productivity of the major parts of economy, the GDP grows with a much smaller rate.} \label{figgrowthes}
\end{figure}

I ran a simple code to simulate conditions and measure growth rates. I supposed that productivity grew $\%5$ annually in sector $A$. Also I supposed that saving was with a rate so how that at each year the rate of return on capital was equal to $5.5\%$. In other words I ignored the accumulation of capital, and supposed that in each year, given the new technology, capital is in a level that the rate of return keeps its sustainable value. 
Given this rate, we could find physical capital per labor in each sector through Eq. (\ref{capitalab}) and then find the production in each sector for both islands. The wage was $200\$$ in 1900 and faced a $6\%$ growth annually. Part of the wages growth was real and part of it was inflation. Calculations concerning real GDP revealed the amount of inflation. Results concerning real growth are graphed in Fig. \ref{figgrowthes}.

In each graph we have two measurements for the real growth rate. In one of the measurements to calculate the real GDP I have taken the prices of year 1900 as the base price and in the other one for each year I have taken its previous year as the base year. As it can be seen not only each measurement suggests a different rate for the growth but also none of the measurements suggest a $5\%$ growth rate. When for each year we consider the previous year as the base year, we observe that as time goes by our measurement suggests a smaller and smaller growth rate. After 98 years where productivity in sector A still experiences a $5\%$ growth, GDP grows by $2.5\%$ in the North Island while it grows only as much as $0.9\%$ in the South Island. The average growth for this  98 year period is $3.0\%$ for the North Island and $1.6\%$ for South Island. So, though productivity in sector A grows 119 folds in about a century, if for each year we consider the previous year as the base year, then based on the central banks report, the real GDP has grown 18 folds in the North Island and only 5 folds in the South Island.

\begin{figure}
        \centering
        \begin{subfigure}[b]{0.5\textwidth}
                \centering
                \includegraphics[width=\textwidth]{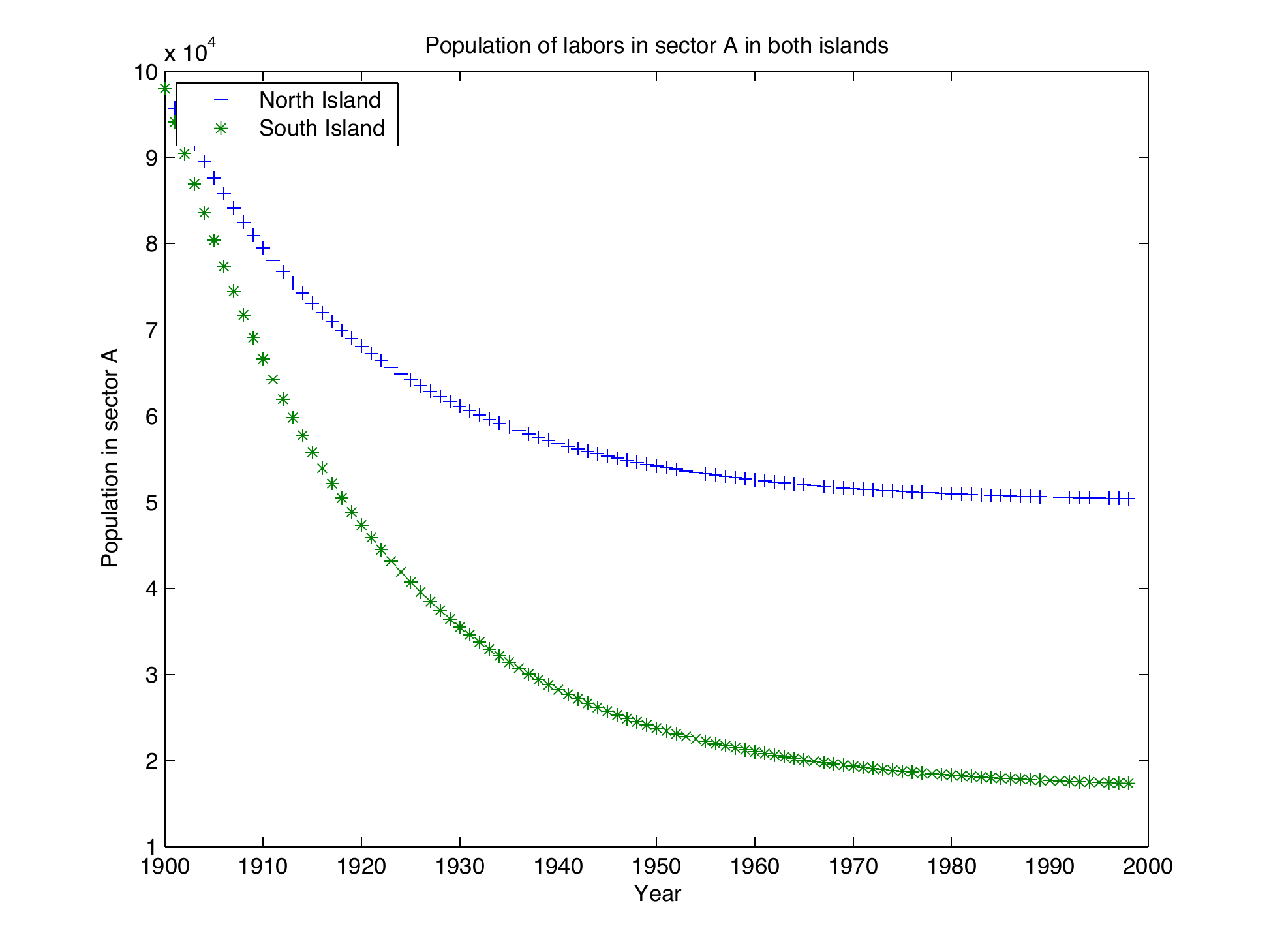}
                \caption{}
                \label{fig:labor}
        \end{subfigure}%
         %add desired spacing between images, e. g. ~, \quad, \qquad etc.
        \begin{subfigure}[b]{0.5\textwidth}
                \centering
                \includegraphics[width=\textwidth]{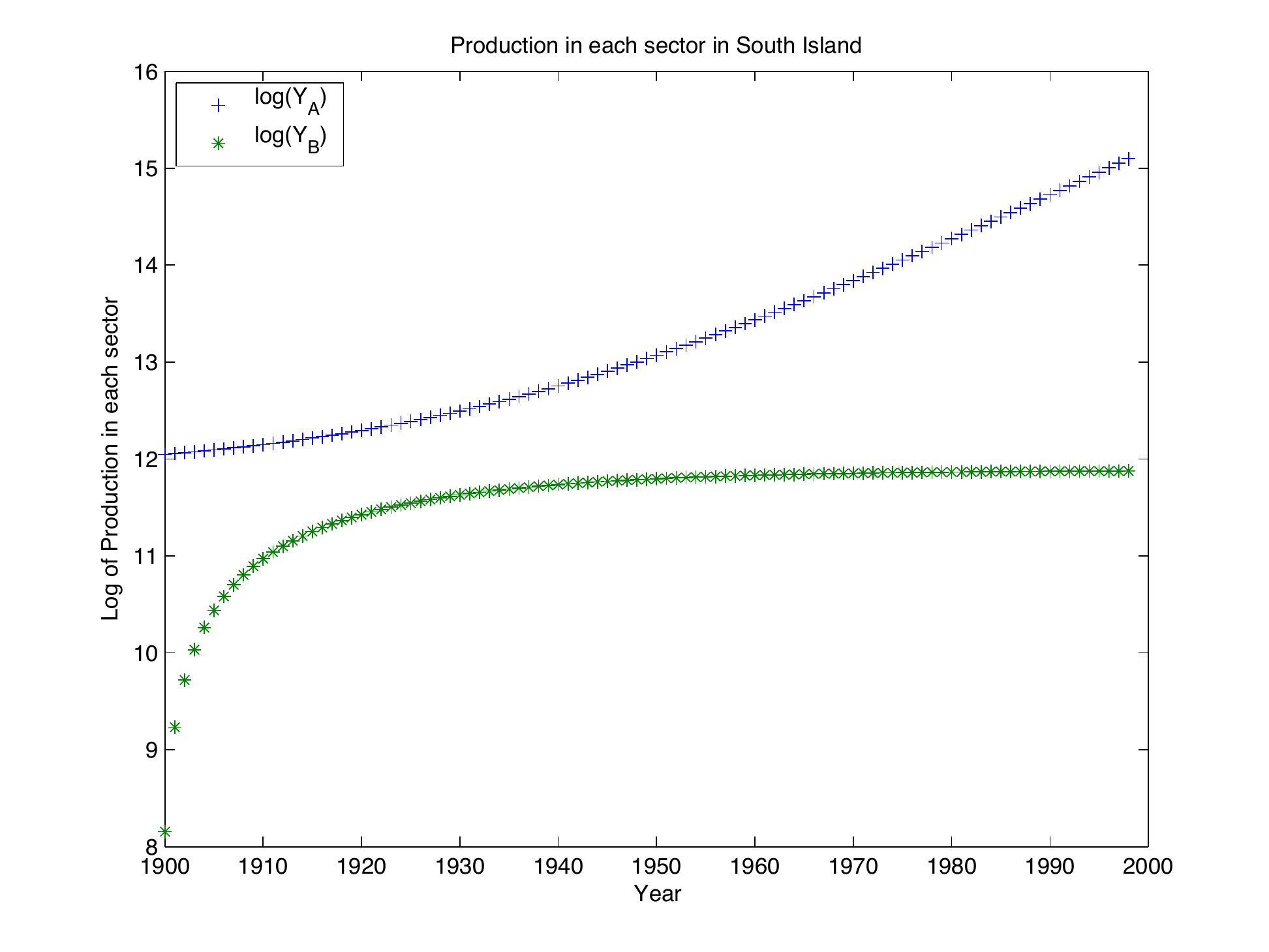}
                \caption{}
                \label{fig:siproduct}
        \end{subfigure}

     %add desired spacing between images, e. g. ~, \quad, \qquad etc.     
        \begin{subfigure}[b]{0.5\textwidth}
                \centering
                \includegraphics[width=\textwidth]{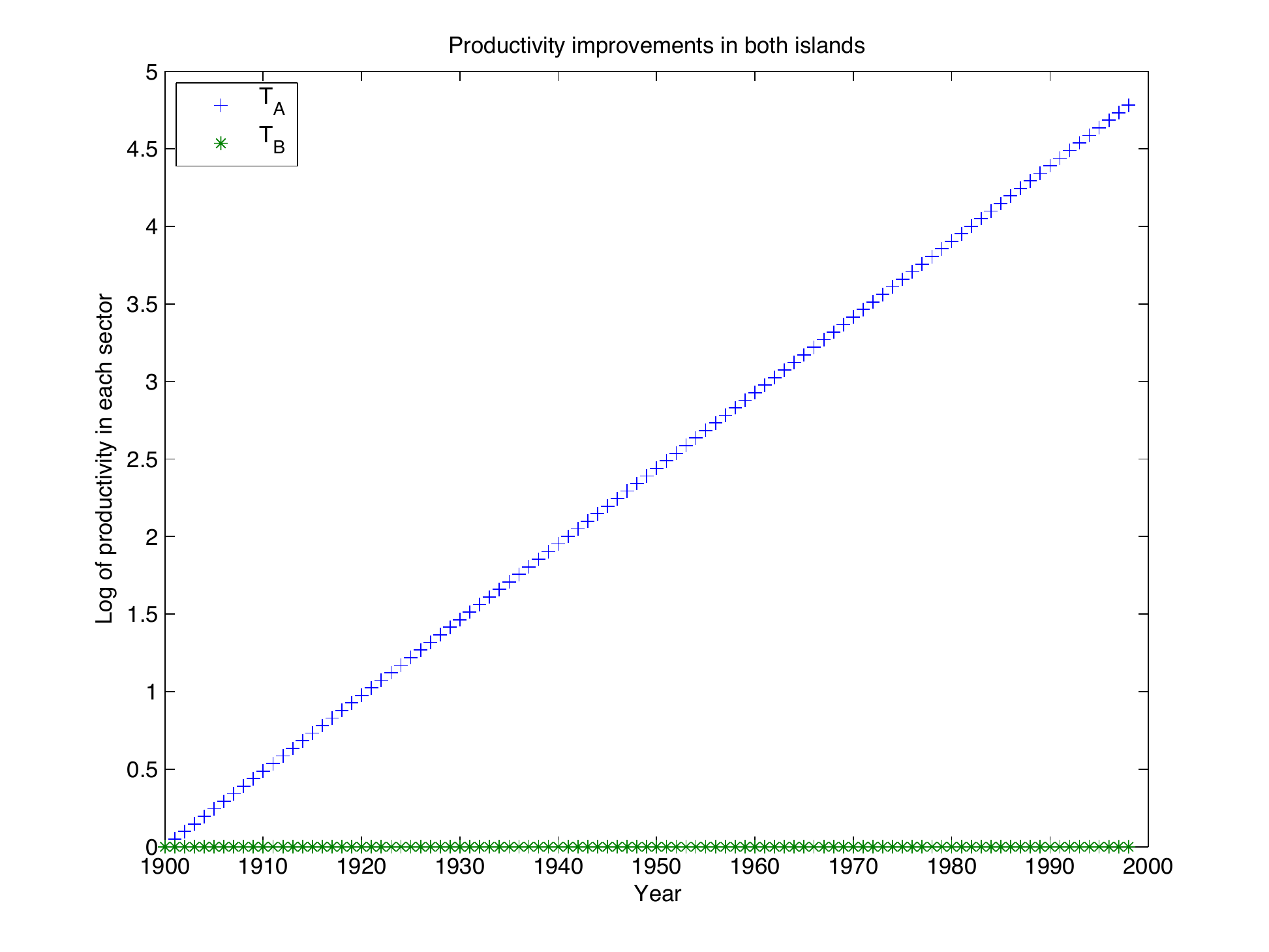}
                \caption{}
                \label{fig:gukll}
        \end{subfigure}%
              %add desired spacing between images, e. g. ~, \quad, \qquad etc.     
        \begin{subfigure}[b]{0.5\textwidth}
                \centering
                \includegraphics[width=\textwidth]{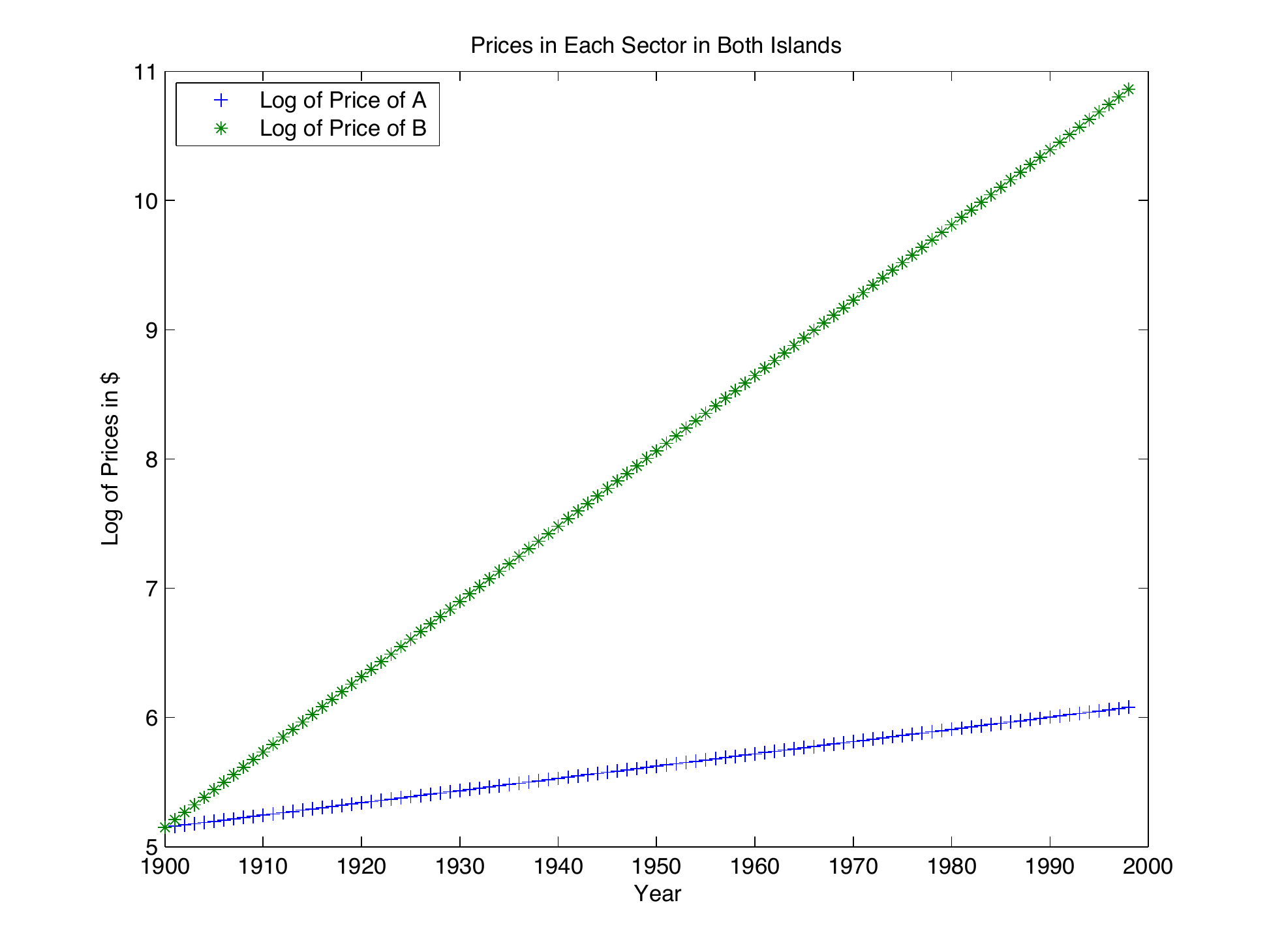}
                \caption{}
                \label{fig:siprices}
        \end{subfigure}%

        \caption[font=footnotesize]{ a) As long the productivity increases in sector A, and the minimum desire for food is fulfilled, population of labors in this sector decreases in both Islands. The final shape of distribution of labors however depends seriously on the utility indifference curves. b)Despite the decrease in the number of labors of sector A, the production in this sector is always increasing. c) Despite the constant productivity in sector B, in sector A it increases by $5\%$ annually. d) Because of the different rate of advancement in productivity in each sector, we face different rate of inflation for each product. } \label{figlaborsand}
\end{figure}

Three phenomena contribute towards obtaining these results. First of all, the utility preference makes big influence. When productivity is small, $98\%$ of labors work in sector A provide the minimum of food we need which is $N_0$. As long as the productivity is increasing the minimum desire for food is fulfilled. Now, people's appetite for more food relaxes and thereby some efforts would be put on the sector B which is a luxury sector. So, though productivity in sector A which at the beginning of the period compensates for $98\%$ of the GDP grows substantially, GDP does not grow as much as that. This is due to the fact that people do not need so much food. In this case as long as the productivity grows in sector A, its portion of GDP shrinks and thereby its effect on the growth rate shrinks as well. The speed in which the portion of sector A shrinks highly depends on the utility preferences. Figure {\ref{fig:labor}} depicts population of the labors which work in sector A. As it can be seen, while in the North Island in the final stages, the economy devotes $50\%$ percent of its labors to the sector A, where in the South Island only $17\%$ percent is devoted to this sector. In this case productivity improvements in sector A have a higher effect in the North Island compared with the South Island.

Reallocation of labor as a result of improvement in technology or what is called "structural changes" is one of the early consequences and predications of the Baumol's cost disease.  According to Alston et. al. 2010, in 1869 $37\%$ of Americans were working in the agriculture sector, where in 2006 only $0.8\%$ had dedicated themselves to that sector. This reallocation happens in all sectors. If we think of automotive industry we find that people are not willing to have more than one or two cars in average. Now, for a closed economy as long as car shortage exists, any improvement in productivity would directly affect the number of produced cars in addition to the  GDP growth rate. When the market is saturated, however things are different. In this case the number of cars produced annually changes hardly and improvement in productivity would not result in a growth of production in this sector. The effect of productivity improvements will be however indirect. In this case as a result of improvement in productivity, some labors and capital are then free to leave the automotive industry and go to some new sectors. The quantitative effect of migration of labors to GDP growth rate will be addressed later in this note. \footnote{When the market is saturated, then improvement in productivity may result in qualitative improvements of products which is out of the scope of the present work.}

If we look at the equations of the previous section we observe that reallocation of labor is not the only reason why the effect of growth rate in sector A gradually tones down. Actually in South Island for example after 70 years the market reaches its saturation level for allocation of labors. This resulted in the number of labors in sector A to be one fifth of sector B (see Fig. {\ref{fig:labor}}). In spite of the small population of labors in sector A due to productivity improvements, production in sector A in 1971 was still 6 folds more than sector B. At the beginning of the period, prices of both products were the same. If in 1971 prices were still equal, then still the main portion of GDP would belong to sector A. In spite of this picture, each product experiences a different rate of inflation (See Fig. {\ref{fig:siprices}}). As a result, given the new prices, the portion of sector A in the GDP is much smaller in this year. So, if the central bank sets the 1970 prices as the base price for measuring the real GDP of 1970 and 1971, it will report a growth rate as of $1.0\%$. If this bank however in 1970 and 1971 still measures the real GDP based on prices of 1900 then the growth rate will be about {3.8\%}. So, beside utility preferences which addresses reallocation of capital and labor, different inflation rates have a serious impact on our measurements for the GDP growth rate. As long as the prices experience different inflation rates, report on the GDP growth rates becomes highly dependent on the base year. In Figure {\ref{figgrowthes}} we observe that in the early years, if we choose year 1900 as the base year, our estimation for the growth rate will be smaller than that of the other measurement. Later however if for each year we choose the last year as the base year, our measurement will be quite smaller. In Figure \ref{figgrowthess} I have provided four measurements for the growth rate in the South Island. The curve indicated by the "+" sign is for the measured real GDP growth of two subsequent years, where the base year for the prices was $15$ years prior to them. As we expect each measurement itself suggests a different rate for the real growth. Another interesting observation is that by selecting the year 1900 as the base year, we would see an average of $\%5$ for the GDP growth rate despite the cost disease. But practically, selecting a old dates for the base year creates difficulties which disables measurements. 

 As it can be seen in my toy model, in the long run different base years may suggest different growth rates. Now, if one, as an stylized fact, claims that GDP sustains a constant growth rate over time, she should make it clear which measurement of the growth rate is implemented. The author believes that only a measurement that considers a recent year as the base year suggests such a fact. If for example for each year we could set 1900 as the base year, though we would have problem on the quality of improvements as well as problems with the new goods, but we would observe however that the growth rate no longer sustains a constant rate in the long run. I will provide later in this note our justification why the current measurements of the central bank which regularly chooses a recent year as the base year, addresses a constant growth rate as a stylized fact. So, based on this simulation I conclude that when we want to discuss the origins of the stylized facts we need to consider a couple of factors. The first influential factor is our utility preferences which can substantially affect the growth rate and allocation of capital and labor. Another important factor is heterogeneous improvements which results in heterogeneous inflation rate and thereby vulnerable measurements.

%%%%%%%%%%%%%%%%%%%%%%%%%%%%%%%%%%%%%%%%%%%%%%%%%%%%%%%%%%%%%%%%%%%%%%
%%%%%%%%%%%%%%%                                                                                     %%%%%%%%%%%%%%%
%%%%%%%%%%%%%%%             \section{A misperception of GDP growth rate}
%%%%%%%%%%%%%%%                                                                                     %%%%%%%%%%%%%%%
%%%%%%%%%%%%%%%%%%%%%%%%%%%%%%%%%%%%%%%%%%%%%%%%%%%%%%%%%%%%%%%%%%%%%%

\subsection{A Model to Violate All Properties}

In this section I provide a toy model in which all three properties can be violated. Concerning property $1$ in this experiment we improve the productivity for both sectors of the market $S$ folds. Surprisingly we observe that our measurement for the GDP growth rate subject to different cases overestimates/underestimates the real growth rate. Concerning property $3$ we observe that the central banks of two countries with exactly the same initial and final conditions report completely different real growth rates for the average of the period. 

Consider three Islands which I call North, Middle and South Islands. In this experiment I suppose that for all of them we have $\Omega=5$ and $N^0=1.6711$.  Other initial conditions are as the initial conditions of the experiment provided in the previous section. In this case, based on Eq. (\ref{priceintermediate}) we have
\bea
P_A^{1900}=   172.34\$\;\;\;\;\;\;\;\;\;\;\;P_B^{1900}=   172.34\$.
\eea
As of the last simulation, given these initial conditions, $98\%$ of labors work in sector A, where in 1900 the production has been $Y_A^{1900}=176000$ and $Y_B^{1900}=3481$. So the GDP in all Islands at the beginning is $30,000,000\$$.

Now, let's suppose that during a 98 year period starting from 1900, productivity in both sectors grows around 19 folds. In other words while in 1900 we have $T_A=T_B=1$ in 1998 we have $T_A=T_B=18.93$. So, since in both sectors productivity has grown by 19 folds, based on the central bank report we percept a growth as of 19 folds for real GDP. Now, surprisingly we observe that central banks may percept and report completely different results in different countries. Though initial and final level of production in three countries are the same, but when they follow different patterns of growth, central banks percept completely different rates of real growth for the whole period.
 
Initial and final conditions in these three islands is exactly the same. The pattern of growth is however different in each country. In the Middle Island, $R\&D$ projects are devoted to both sectors equally and we have a smooth $3.0\%$ growth in productivity. In other words for the $i_{th}$ year we have
\bea
T^{i}_A=(1.0305)*T_A^{i-1}\;\;\;\;\;\;\;\;\;\;\;\;\;\;\;\;\;T^i_B=(1.0305)*T_B^{i-1}
\eea
Since technological advancements are exactly the same for both sectors, they will experience an equal rate of inflation. 

Now, I suppose that in the North Island, the NSF of the country supports researches on agriculture at early years and as a result the productivity in sector A grows very fast at the beginning. Later more and more researches is devoted to the sector B. So I suppose a pattern of growth in productivity in each sector as 
\bea
T^{1900+i}_A=(1+0.06*\frac{100-i}{99})*T_A^{1900+i-1}\;\;\;\;\;\;\;\;T^{1900+i}_B=(1+0.06*\frac{i+1}{99})*T_B^{1900+i-1},
\eea
in which $i$ ranges from 1 to 98. Figure {\ref{fig:niproductivity2}} depicts the pattern of this growth. As it can be seen from this equation and the figure, in the end, productivity in both sectors grows 19 folds. But each sector follows a different pattern. 

In the South Island, NSF follows a different strategy and prefers to first support researches in sector B. So, in contrast to the North Island, the productivity for the South Island grows as
\bea
T^{1900+i}_A=(1+0.06*\frac{i+1}{99})*T_A^{1900+i-1}\;\;\;\;\;\;\;\;T^{1900+i}_B=(1+0.06*\frac{100-i}{99})*T_B^{1900+i-1}.
\eea
Now, we have three countries with exactly the same initial and final productivities and exactly the same initial and final GDP. In 1998 in all three islands we have $Y_A=690852$ and $Y_B=2604761$. In the middle age however pattern of the growth has been different for each country, and thereby allocation of labors and rate of inflation for each sector in each country is different. Figure \ref{fig:2gdpgrowthrate} depicts the real growth rate for each country reported by the central banks. For this growth rate, the central banks for each year consider its previous year as the base year. Figure \ref{fig2gdpavereage} shows the mean growth rate for each Island. As it can be seen, surprisingly the average for real growth rate is quite different in each country. Based on annual reports, one concludes that the average of growth for the whole period has been $3.1\%$ in the Middle Island which is what we expect. In the North Island however the average of the growth rate after 98 years is $3.6\%$. In the South Island things are even more surprising since the average of growth is $2.2\%$. So, based on the central banks reports one can conclude that after 98 years, the real GDP has grown 19 folds in the Middle Island, 31 folds in the North Island and 8 folds in the South Island. All of these rates are what we percept which is seriously affected by utility preferences and the growth pattern in different sectors. Now, though initial and final conditions are exactly the same, the governing parties should be proud in the North Island and ashamed in the South Island. These statements in the context of real world could be interpreted as; if every thing e.g. number of cars, houses, and motor cycles, etc. increases 19 folds in a hundred years, we expect a $3$ percent growth rate, while the central bank would report an average growth of $2$ percent.  

If to measure real growth rate for each year, instead of the previous year we set 1900 as the base year, then things will be different. In this case the average growth in all three Islands merge together in 1998 as it had really happened for GDP (See Fig. \ref{fig:2gdpaverage1900}). So, if we can keep the base year constant for a long time, what we percept as the growth rate will be quite different. In this case the governing parties of the three islands have the same situation as they need too.Needless to note that practically choosing a very old initial year for the base year is impossible.

%%%%%%%%%%%%%%%%%%%%%%%%%%%%%%%%%%%%%%%%%%%%%%%%%%%%%%%%%%%%%%%%%%%%%%
%%%%%%%%%%%%%%%                                                                                     %%%%%%%%%%%%%%%
%%%%%%%%%%%%%%%             \section{A misperception of GDP growth rate}
%%%%%%%%%%%%%%%                                                                                     %%%%%%%%%%%%%%%
%%%%%%%%%%%%%%%%%%%%%%%%%%%%%%%%%%%%%%%%%%%%%%%%%%%%%%%%%%%%%%%%%%%%%%

\section{Analysis and Conclusion}

%%%%%%%%%%%%%%%%%%%%%%%%%%%%%%%%%%%%%%%%%%%%%%%%%%%%%%%%%%%%%%%%%%%%%
%%%%%%%%%%%%%%%%%%%%%%%%%%%%%%%%%%%%%%%%%%%%%%%%%%%%%%%%%%%%%%%%%%%%%%
%%%%%%%%%%%%%%%                                                                                     %%%%%%%%%%%%%%%
%%%%%%%%%%%%%%%             \subsection{Analysis }
%%%%%%%%%%%%%%%                                                                                     %%%%%%%%%%%%%%%
%%%%%%%%%%%%%%%%%%%%%%%%%%%%%%%%%%%%%%%%%%%%%%%%%%%%%%%%%%%%%%%%%%%%%%

\subsection{Analytic Investigation}

One may criticize that if we had smooth and continuous growths instead of sharp growths then in experiments we could not precept violation of properties mentioned in the last subsection. In other words one may suggest that these paradoxical conclusions are resulted from discrete measurements. To exclude such possibility I try to have some analytic calculations. 

Consider an economy with a wide range of productions. For GDP of the $i_{th}$ year in the price base of the $j_{th}$ year we have
\bea
GDP_i^j=\Sigma_{a}Y_a^iP_a^j.
\eea 
Now for the growth rate of the $i_{th}$ year in the price base of the $j_{th}$ year we have
\bea
g_i^j=\frac{GDP_i^j-GDP_{i-1}^j}{GDP_{i-1}^j}=\frac{\Sigma_{a}Y_a^iP_a^j-\Sigma_{a}Y_a^{i-1}P_a^j}{\Sigma_{a}Y_a^{i-1}P_a^j}.
\eea
If we want to calculate this growth rate in the price base of the $k_{th}$ year we have
\bea
g_i^k=\frac{GDP_i^k-GDP_{i-1}^k}{GDP_{i-1}^k}=\frac{\Sigma_{a}Y_a^iP_a^k-\Sigma_{a}Y_a^{i-1}P_a^k}{\Sigma_{a}Y_a^{i-1}P_a^k}.
\eea
Now, if we for all production have a unique inflation rate then, we have
\bea
P_j^k=\Pi_{m=j+1}^{k}(1+r_m)P_i^j,
\eea
in which $r_m$ is inflation in the $m_{th}$ year, which is the same for all productions in all sectors. Now for the growth rate we have
\bea\begin{split}
&g_i^k=\frac{\Sigma_{a}Y_a^iP_a^k-\Sigma_{a}Y_a^{i-1}P_a^k}{\Sigma_{a}Y_a^{i-1}P_a^k}
\cr& =\frac{\Sigma_{a}Y_a^i\Pi_{m=j+1}^{k}(1+r_m) P_a^j-\Sigma_{a}Y_a^{i-1} \Pi_{m=j+1}^{k}(1+r_m)P_a^j}{\Sigma_{a}Y_a^{i-1}\Pi_{m=j+1}^{k}(1+r_m) P_a^j}\cr&=\frac{\Pi_{m=j+1}^{k}(1+r_m)[\Sigma_{a}Y_a^iP_a^j-\Sigma_{a}Y_a^{i-1}P_a^j]}{\Pi_{m=j+1}^{k}(1+r_m)[\Sigma_{a}Y_a^{i-1}P_a^j]}=g_i^j.
\end{split}\eea
As it can be seen, if inflation for all productions were the same, then the growth rate would be independent of the base year and violation of properties II and III would not happen any more. Violation of property I however could result from the reallocation of labors and capital and may always survive. 

Now let's consider a smooth growth. For a time interval of $T$ we divide it to infinitesimal time intervals $dt$. At any time $t$ the infinitesimal growth rate can be defined as
\bea
dg=\frac{\Sigma_aY_a(t)P_a(t-dt)-\Sigma_aY_a(t-dt)P_a(t-dt)}{\Sigma_aY_a(t-dt)P_a(t-dt)}=\frac{\Sigma_a\partial_tY_a(t-dt)dtP_a(t-dt)}{\Sigma_aY_a(t-dt)P_a(t-dt)},
\eea
in which $Y_a(t)$ and $P_a(t)$ are production and price of goods $a$ at a given time $t$. Now, to examine the whole growth we may expect
\bea\begin{split}
&\int_0^Tdg=\int_0^T\frac{\Sigma_a\partial_tY_a(t-dt)P_a(t-dt)}{\Sigma_aY_a(t-dt)P_a(t-dt)}dt=\int_0^T\frac{\Sigma_a\partial_tY_a(t)P_a(t)}{\Sigma_aY_a(t)P_a(t)}dt\cr&=\int_0^T\frac{\partial\log{(\Sigma_aY_a(t)P_a(t))}}{\partial t}dt-\int_0^T\frac{\Sigma_aY_a(t)\partial_tP_a(t)}{\Sigma_aY_a(t)P_a(t)}dt\cr&=log(\frac{\Sigma_aY_a(T)P_a(T)}{\Sigma_aY_a(0)P_a(0)})-\int_0^T\frac{\Sigma_aY_a(t)\partial_tP_a(t)}{\Sigma_aY_a(t)P_a(t)}dt,
\end{split}\eea
which results in
\bea
log(\frac{GDP(T)}{GDP(0)})=\int_0^Tdg+\int_0^T\frac{\Sigma_aY_a(t)\partial_tP_a(t)}{\Sigma_aY_a(t)P_a(t)}dt.
\eea

In this equation in the left, we have the logarithm of nominal GDP. In the right, the first term is real growth in the whole period, and the second term is GDP deflater. Now let's consider three economies with similar initial and final conditions as of experiment $II$. As it can be seen in Fig. \ref{fig:productivity}, if we look at the space of technologies, though initial and final conditions are the same, the pattern of growth in interval time is different for each country. In this part of the work I have left capital accumulation aside and have supposed that at each term after the innovation in technology the production function is modified as well to its stable point. In this case in the last equation, the left side is the same for all three economies. In the right side, the first term is the total growth measured by central banks and the second term is GDP deflater. If we want to observe that all three banks report a total growth equal to each other in all three countries, then we need the total measurement for GDP deflation to be the same in every country. So, to eliminate the paradox observed in experiment $II$ we need that for each economy subject to have initial and final conditions, the second term in the right side to be independent of the path. Let's dig this fact a bit more. For total measure of GDP deflater in the last equation we notice
\bea
\int_0^T\frac{\Sigma_aY_a(t)\partial_tP_a(t)}{\Sigma_bY_b(t)P_b(t)}dt=\int_0^T\frac{\Sigma_aY_a(t)\Sigma_c\frac{\partial P_a(t)}{\partial T_c}\frac{\partial T_c}{\partial t}}{\Sigma_bY_b(t)P_b(t)}dt=\int_0^T\Sigma_c\frac{\Sigma_aY_a\frac{\partial P_a}{\partial T_c}}{\Sigma_bY_bP_b}dT_c.
\eea
Now, to have this integral to be independent of any path; exactly similar to a conservative force; we need 
\bea
\frac{\partial}{\partial T_d}(\frac{\Sigma_aY_a\frac{\partial P_a}{\partial T_c}}{\Sigma_bY_bP_b})=\frac{\partial}{\partial T_c}(\frac{\Sigma_aY_a\frac{\partial P_a}{\partial T_d}}{\Sigma_bY_bP_b}).
\eea
Keeping nominal wage constant over time and plugging prices from Eq. (\ref{priceintermediate}) in this equation we find
\bea\label{bala}
\frac{\partial}{\partial T_d}(-\frac{L_cW/\lambda_cT_c}{\Sigma_bY_b(t)P_b(t)})=\frac{\partial}{\partial T_c}(-\frac{L_dW/\lambda_dT_d}{\Sigma_bY_b(t)P_b(t)}).
\eea
If we want our measurement to be independent of the growth path we then need the Eq. (\ref{bala}) equation to hold for all choices of $d$ and $c$. For out bi-sector economy we need
\bea
\frac{\partial}{\partial T_A}(\frac{L_B}{\lambda_BT_B\Sigma_bY_bP_b})=\frac{\partial}{\partial T_B}(\frac{L_A}{\lambda_AT_A\Sigma_bY_bP_b}).
\eea
A trivial calculation through plugging variables from Eq. (\ref{labormispersepction}) and equations in section (\ref{secanomalus}) reveals that the last equation does not hold. So, analytic calculation supports that measurement for GDP growth rate can violate invariance. 
%%%%%%%%%%%%%%%%%%%%%%%%%%%%%%%%%%%%%%%%%%%%%%%%%%%%%%%%%%%%%%%%%%%%%
%%%%%%%%%%%%%%%%%%%%%%%%%%%%%%%%%%%%%%%%%%%%%%%%%%%%%%%%%%%%%%%%%%%%%%
%%%%%%%%%%%%%%%                                                                                     %%%%%%%%%%%%%%%
%%%%%%%%%%%%%%%             \subsection{Analysis }
%%%%%%%%%%%%%%%                                                                                     %%%%%%%%%%%%%%%
%%%%%%%%%%%%%%%%%%%%%%%%%%%%%%%%%%%%%%%%%%%%%%%%%%%%%%%%%%%%%%%%%%%%%%

\subsection{Sustainable Growth Rate; a Result of Measurements}

In this section I aim to prove that if Kaldor's 5 other facts is fulfilled and beside saving rate is around a sustainable rate, then the sixth fact emerges as  a result of measurements. Let's suppose that facts 1 and 2 are held. Facts 2 reads as
\bea
LS=\frac{L_tW}{Q_t}=\frac{Q_t-K_t(R_c+\bar\delta)}{Q_t}=1-\frac{K_t(R_c+\bar\delta)}{Q_t}.
\eea
Now, capital output ratio is
\bea\label{kaldrocapital}
\frac{K_t}{Q_t}=\frac{1-LS}{R_c+\bar\delta}.
\eea
Now if beside labor share (LS) and rate of return, average of depreciation is sustainable, then capital output ratio is sustainable too. Let's call this rate $C_K$ or $C_K=K_t/Q_t$. Let's suppose that in a country all Kaldor's facts except the fact 6 are held. Now, for the coming year we find new technologies in some sectors. For these sectors $T_a\rightarrow T^{\prime}_a$. According to Eq. (\ref{capitallevel5}) physical capital needed for this new level of production directly depends on $T_a$. If $T_a$ grows, then to trigger production to the new level we need to invest more capital. Physical capital needed for new investment is
\bea
\Delta N_a=(\frac{1-\lambda_a}{R_c+\delta_a})^{\frac{1}{\lambda_a}} G_a T_a,
\eea
in which $G_a$ is growth rate in technology in sector $a$. Money value capital needed for this improvement is
\bea\label{capitalsustainablee}
\Delta K_a=\Delta N_a P_a=G_a K_a
\eea
so for aggregate capital we have
\bea
\bar \Delta K_t=\bar G K_t.
\eea
Now, the question is that how long it takes that we invest a money value capital equal to $\bar G K_t$.
Time period needed for this investment is
\bea
\Delta t=\frac{\bar G K_t}{SQ_t}
\eea
in which $S$ is national saving rate. Now, based on Eq. (\ref{kaldrocapital}), it will be equal to
\bea
\Delta t=\frac{G(1-LS)}{S(R_c+\bar\delta)}
\eea
During this time period since capital grows proportional to $\bar G$, then as a consequence of fact 2, then labor share grows with this size and thereby the whole economy grows proportional to this rate. The annual growth rate however is
\bea
\bar g=\frac{G}{\Delta t}=\frac{S(R_c+\bar\delta)}{1-LS}
\eea
As it can be seen, average growth rate ($\bar g$) is independent of growth rate for technology ($\bar G$).
As long that labor share, depreciation rate, rate of return on capital, average of depreciation rate, and national saving rates are sustainable, then real growth rate is sustainable too. According to Modigliani life cycle hypothesis
as long that the cycle pattern of life is sustainable then saving rate is sustainable too. GDP growth has fluctuated around its sustainable rate for more than a century. Long before, the main part of economy was agriculture. Then industrial sectors took that position and now everything is turning into service side. R\&D has grown substantially at the beginning of 20s century. How productivity should grow with a sustainable rate? Does by some strange reason technology is discovered by a sustainable rate around \%2? It seems too naive to think that technology is discovered by a sustainable rate with these different movements from farm sector to service sector and with these substantial changes in education, skillfulness and R\&D programs. The author belief is that a long lasting origin such as life-cycle pattern and a matter of measurement should have complied to illustrate such result. Let's dig this problem a bit more.

Economy consists of heterogeneous goods. We have goods such as food which are essential needs and have low elasticity to price. No matter what the level of technology is, the consumption of food fluctuate around a sustainable rate per capita. We may ask now what happens if productivity grows in such sectors. Trivially production for these sectors keeps the same level in a closed economy. Even in open economy, many goods are not movable or hard to move and with or without improvement production for them are kept at the same level. For these sectors improvement in technology could result in lower prices. When we measure real GDP however since we adjust prices of the following years, then despite improvement in technology we do not see any growth in size and thereby growth rate. What happens then? As a result of improvement in technology in these sectors, more labors and capital are free to go to the other sectors. They go to the sectors which provide goods or services which have elasticity to prices. They produce new goods and services. We consume their production however if price for them is reasonable. Price for new sectors is reasonable if and only if they satisfy market clearing condition or
\bea
\Sigma_bP_b\Delta Y_b=\Sigma_b[\Delta L_b W+\Delta K_b (R_c+\delta_b)]
\eea         
in which $b$ stands for sectors which we have new investment, $\Delta L_b$ is new hired labors in these sectors and $\Delta K_b$ is new investment. Aggregate of new investments is about $SQ_t$. Now, since in both following years labor share has been twice capitalists share, then we have
 \bea
\Sigma_b\Delta L_b W=\frac{LS}{1-LS}\Sigma_b\Delta K_b (R_c+\delta_b).
\eea
As a result we find
 \bea
\Sigma_bP_b\Delta Y_b=\frac{1}{1-LS}\Sigma_b\Delta K_b (R_c+\delta_b)=\frac{R_c+\bar\delta}{1-LS}SQ_t.
\eea   
So, the real growth rate will be
\bea
g=\frac{S(R_c+\bar\delta)}{1-LS}.
\eea
This result is independent of production functions. If we look at historical data we see that this picture works. Looking at figure \ref{figgdp} we observe that during the past decades service share in GDP has grown by 50\%. Share of goods has declined almost by the same rate. Despite this allocation of efforts to service side, the real growth rate has kept its historical 2\% rate. It means that our scenario works. Sustainable growth rate more than everything is a result of four facts. The first fact is sustainable rate of return on capital. The second fact is sustainable share of labor income. The Cobb-Douglas aggregate production function can guarantee it.\footnote{In Hosseiny 2015 however evidences have been provided to show that Modigliani life cycle hypothesis can guarantee a sustainable labor share of income.}. The third fact is mobility of labor and capital and equilibrium conditions and the last important fact is our method of measurement.

%%%%%%%%%%%%%%%%%%%%%%%%%%%%%%%%%%%%%%%%%%%%%%%%%%%%%%%%%%%%%%%%%%%%%
%%%%%%%%%%%%%%%%%%%%%%%%%%%%%%%%%%%%%%%%%%%%%%%%%%%%%%%%%%%%%%%%%%%%%%
%%%%%%%%%%%%%%%                                                                                     %%%%%%%%%%%%%%%
%%%%%%%%%%%%%%%             \subsection{Analysis }
%%%%%%%%%%%%%%%                                                                                     %%%%%%%%%%%%%%%
%%%%%%%%%%%%%%%%%%%%%%%%%%%%%%%%%%%%%%%%%%%%%%%%%%%%%%%%%%%%%%%%%%%%%%
\subsection{Central Banks, World Bank and General Relativity}
In recent years report series concerning GDP of different countries and their paradoxical results have been of attention.  Penn World Table and International Comparison Program measure GDP of countries through purchasing power parity (PPP) method. In this measurement prices of different goods are considered the same for all countries. Tradable goods have almost the same price in all countries. Non-tradable goods however have different prices. A PPP method tries to consider the same price for both tradable and non-tradable goods and through aggregate data evaluate GDP of considered countries with the same price base. Through a time series of data we can measure average of real growth rate of listed countries. When we measure this rate we observe that surprisingly the result is different from the ones reported in national accounts. Even once it was observed that China GDP shrunk substantially (see Feenstra et al. 2013-a). Many reasons have been suggested for this problem. A major problem suggested to be choosing a wrong basket of goods for countries. I however emphasize the role of measurement. Prices in PPP method is adjusted to a global set of prices. The problem is that each country has its own soup of firms, its own utility preferences and its own relative prices. In this paper we noticed that measuring GDP for a country itself exposes a paradox. It would be much problematic when you want to compare GDP with prices which may be an outcome of equilibrium conditions of another country. This point has attracted attentions as space-time inconsistency in the literature where by space economists mean different countries. For a fast growing country such as China things will be even more interesting. For this country things are changing fast. Many new brands show up. It makes things to be out of equilibrium. If we accept Modigiliani life cycle hypothesis, then preferences for consumption and saving rate is different in China. So, relative prices should be changing in such country even faster than the other parts of the world. No wonder that problem with China can be huge. It should be added that the effect of indexing in paradoxical results is being considered in current researches beside other problems.\\

{\bf -The gap between China and the US}\\

As the last question let's suppose that we want to know the gap between China and the US and extrapolate the time that economy in China is as big as the US. Is it a simple question even if we suppose that prices are at equilibrium in China? Measuring distance between China and the US is a hard task. The problem is similar to the measurement of some variables in general relativity. China has its own set of equations mentioned in section  II. The US has its own conditions. When China is to pass the gap to the US its meter for GDP will change (see Fig. \ref{figgeneralrelativity}). When per capita GDP grows in China services will grow as well. Different conditions impose different metric for geometry. So if we aim to extrapolate how long does it take for China to pass economy of the US we need to be careful about the metric in the midway. 

Actually we should be careful to use the term metric. An infinitesimal growth in GDP is defined as
\bea
dGDP=\Sigma_aP_adY_a.
\eea
It is more similar to a one-form. So, we don't have a proper definition as of metric. The problems with measurement are however very similar to the same problems in general relativity and thereby we borrowed the term metric - though mathematically untrue. Measurement with GDP growth rate is very similar to the measurement of distance between two points in a manifold. In our paradoxical results in section 3.2 though all 3 countries have an initial and final position they however pass through different paths. Each central bank of these countries measure length of its own path. Central banks at each point consider local equilibrium conditions and local metrics. From the language of general relativity they are different falling observers and their measurements are different from the measurements of World Bank which use its own metric and is an outside observer. In our examples we used the same utility preferences for all countries. People in each country however may have their own preferences. In this case each country has its own geodesic.

%%%%%%%%%%%%%%%%%%%%%%%%%%%%%%%%%%%%%%%%%%%%%%%%%%%%%%%%%%%%%%%%%%%%%
%%%%%%%%%%%%%%%%%%%%%%%%%%%%%%%%%%%%%%%%%%%%%%%%%%%%%%%%%%%%%%%%%%%%%%
%%%%%%%%%%%%%%%                                                                                     %%%%%%%%%%%%%%%
%%%%%%%%%%%%%%%             \subsection{Analysis }
%%%%%%%%%%%%%%%                                                                                     %%%%%%%%%%%%%%%
%%%%%%%%%%%%%%%%%%%%%%%%%%%%%%%%%%%%%%%%%%%%%%%%%%%%%%%%%%%%%%%%%%%%%%
\subsection{Summary}
In this paper I considered invariance of measurements for GDP growth rate. Through studying toy models I observed that dependence of measurements on the pattern of growth can be huge. I then notified that sustainable growth rate is a matter of measurement. By this I mean that if we have Kaldor's five other facts and a sustainable saving rate we then automatically observe a sustainable growth rate. Prices are automatically adjusted to a level that this growth is observed for a central bank which uses local metrics. This observation is subject to business cycles of the market. I then notified that paradoxical results from World Bank reports and from national accounts are similar to the problems for measurement in General Relativity. It should be notified that current equations do not define a Riemannian manifold however.

%---------------------------------------------------------------------
%Bibliography

\captionsetup[figure]{labelfont=bf,textfont={it}}

\begin{figure}
  \centering
    \includegraphics[width=0.5\textwidth]{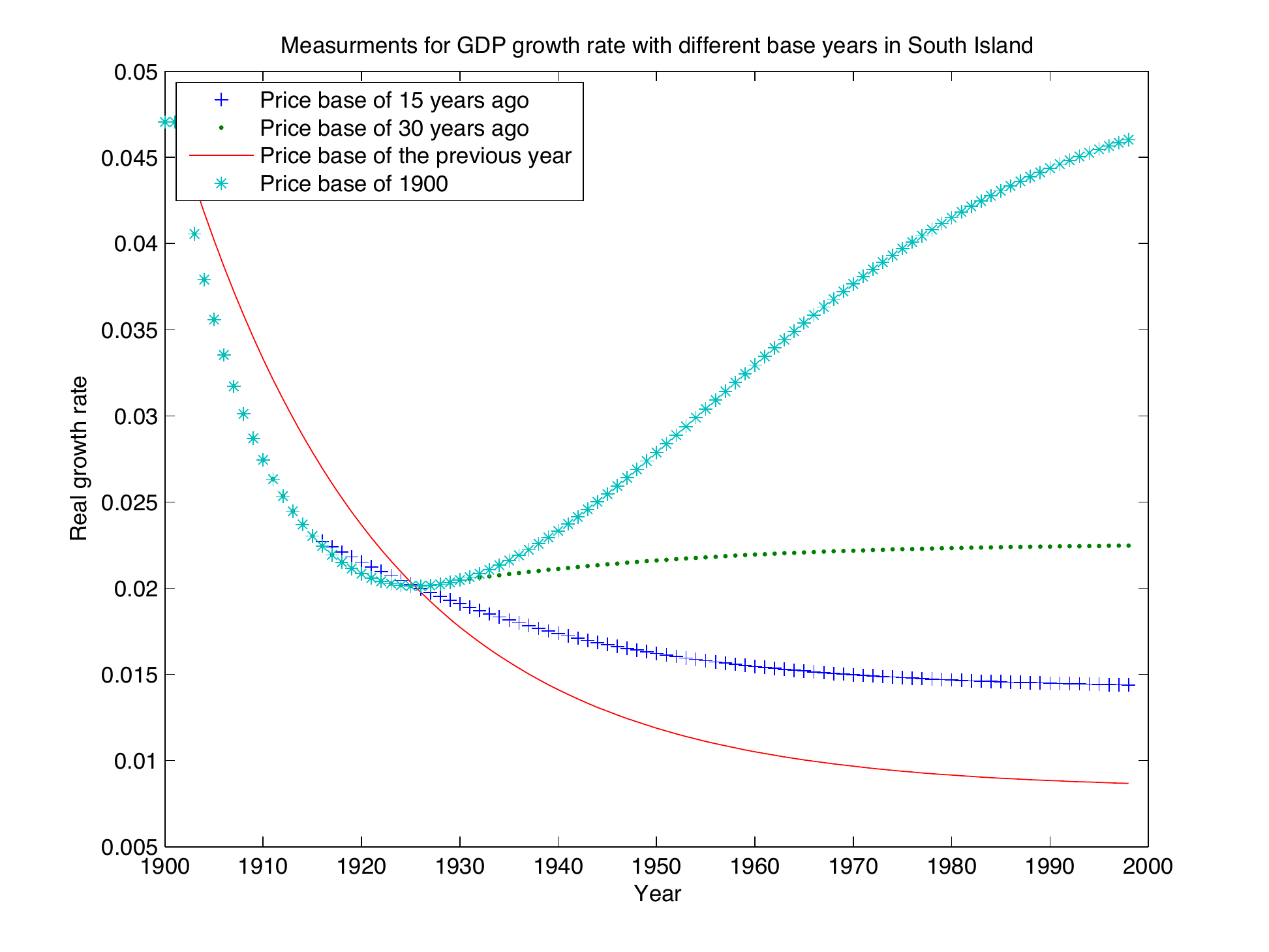}
  \caption{As we set different years as the base year, we obtain different rates of real growth.}\label{figgrowthess}
\end{figure}\label{figgrowthess}

\begin{figure}
        \centering
        \begin{subfigure}[b]{0.4\textwidth}
                \centering
                \includegraphics[width=\textwidth]{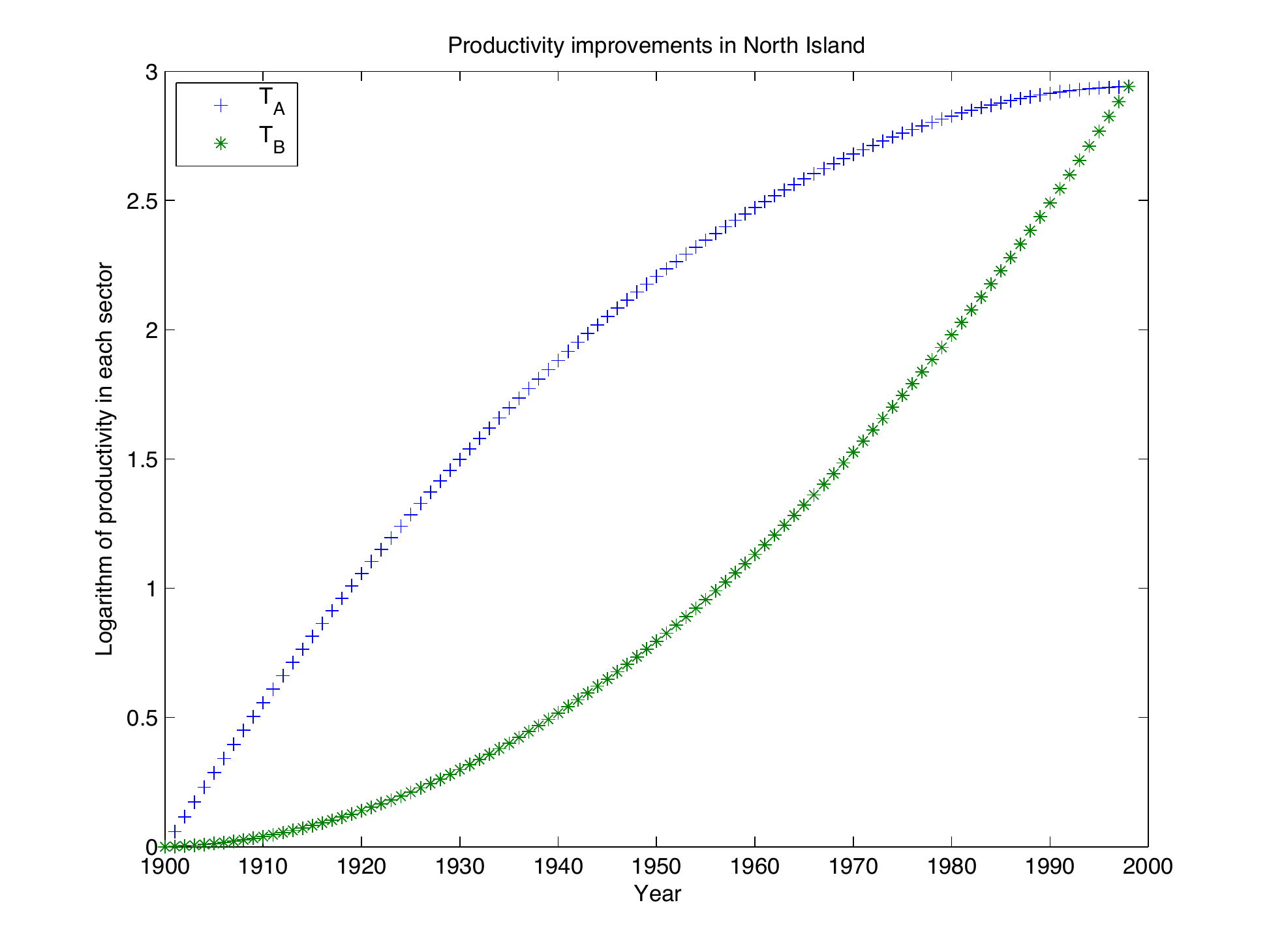}
                \caption{}
                \label{fig:niproductivity2}
        \end{subfigure}%
         %add desired spacing between images, e. g. ~, \quad, \qquad etc.
        \begin{subfigure}[b]{0.4\textwidth}
                \centering
                \includegraphics[width=\textwidth]{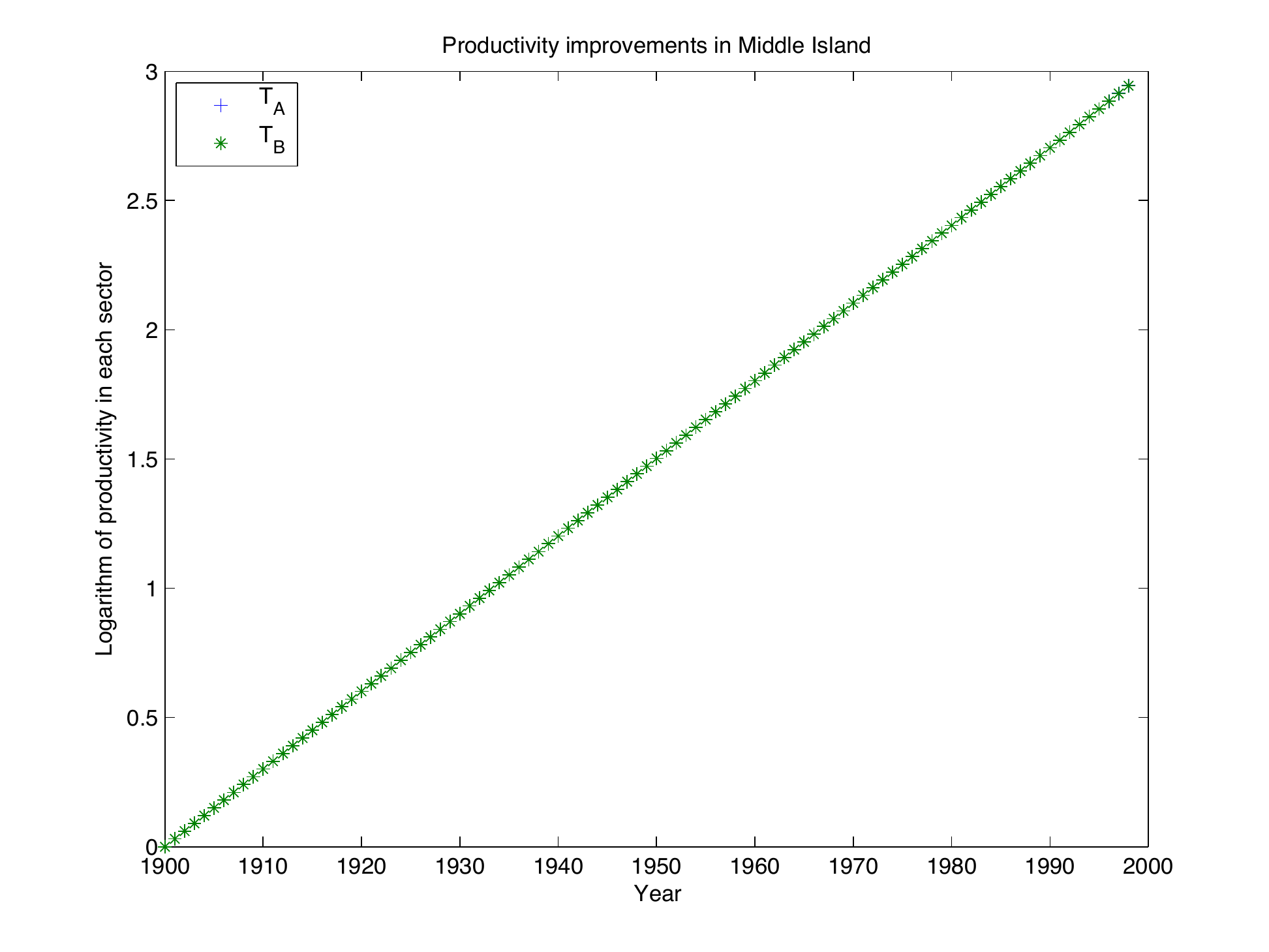}
                \caption{}
                \label{fig:miproductivity2}
        \end{subfigure}
        %add desired spacing between images, e. g. ~, \quad, \qquad etc.
        \begin{subfigure}[b]{0.4\textwidth}
                \centering
                \includegraphics[width=\textwidth]{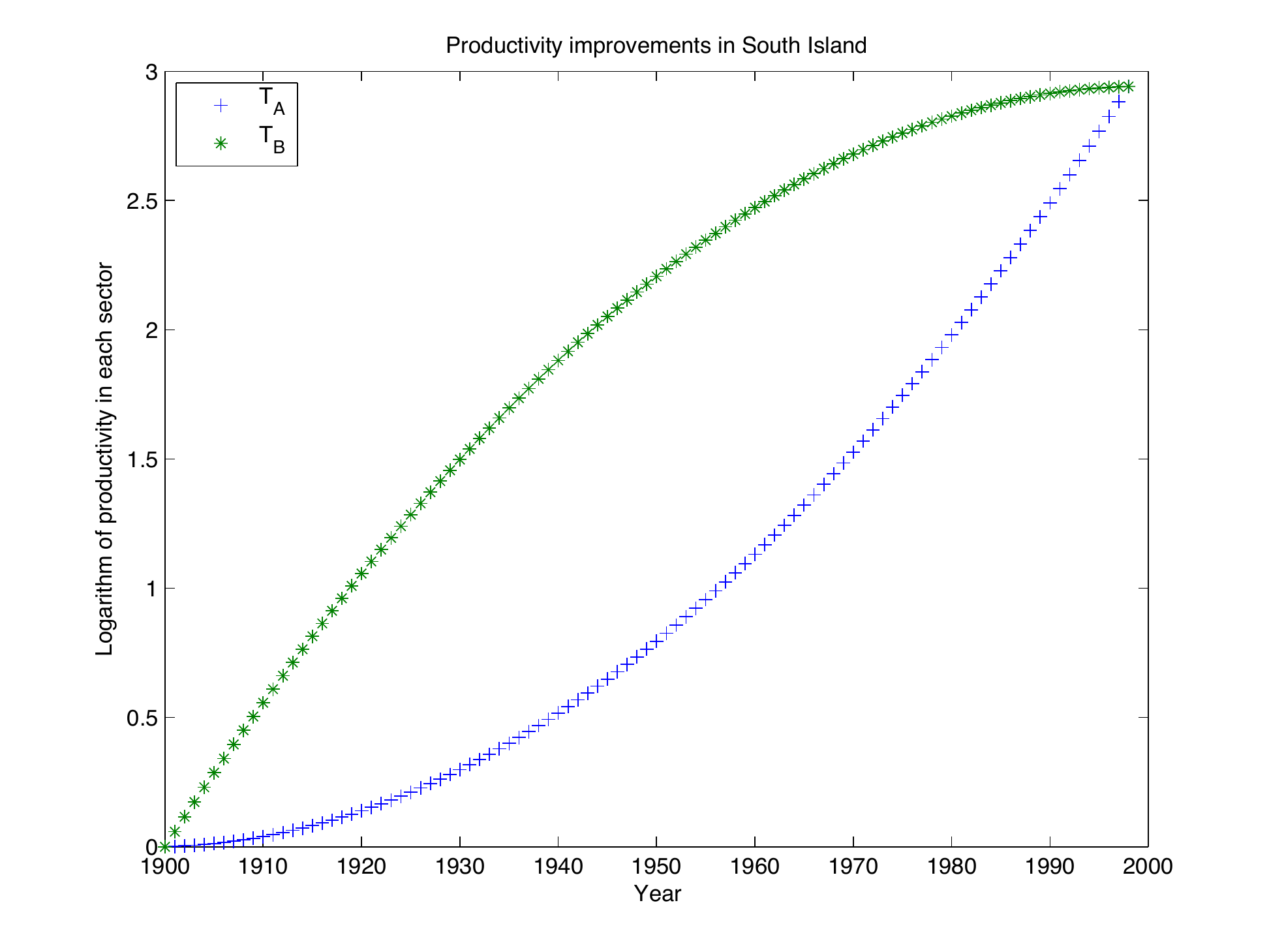}
                \caption{}
                \label{fig:siproductivity2}
        \end{subfigure}

        \caption[font=footnotesize]{Improvement in productivities in a)North Island, b)Middle Island and c)South Islands. After 98 years, the productivity in all three islands stays in the same level. In the middle times however the level of productivity is different in each island.} \label{fig:productivity}
\end{figure}

\begin{figure}
        \centering
        \begin{subfigure}[b]{0.5\textwidth}
                \centering
                \includegraphics[width=\textwidth]{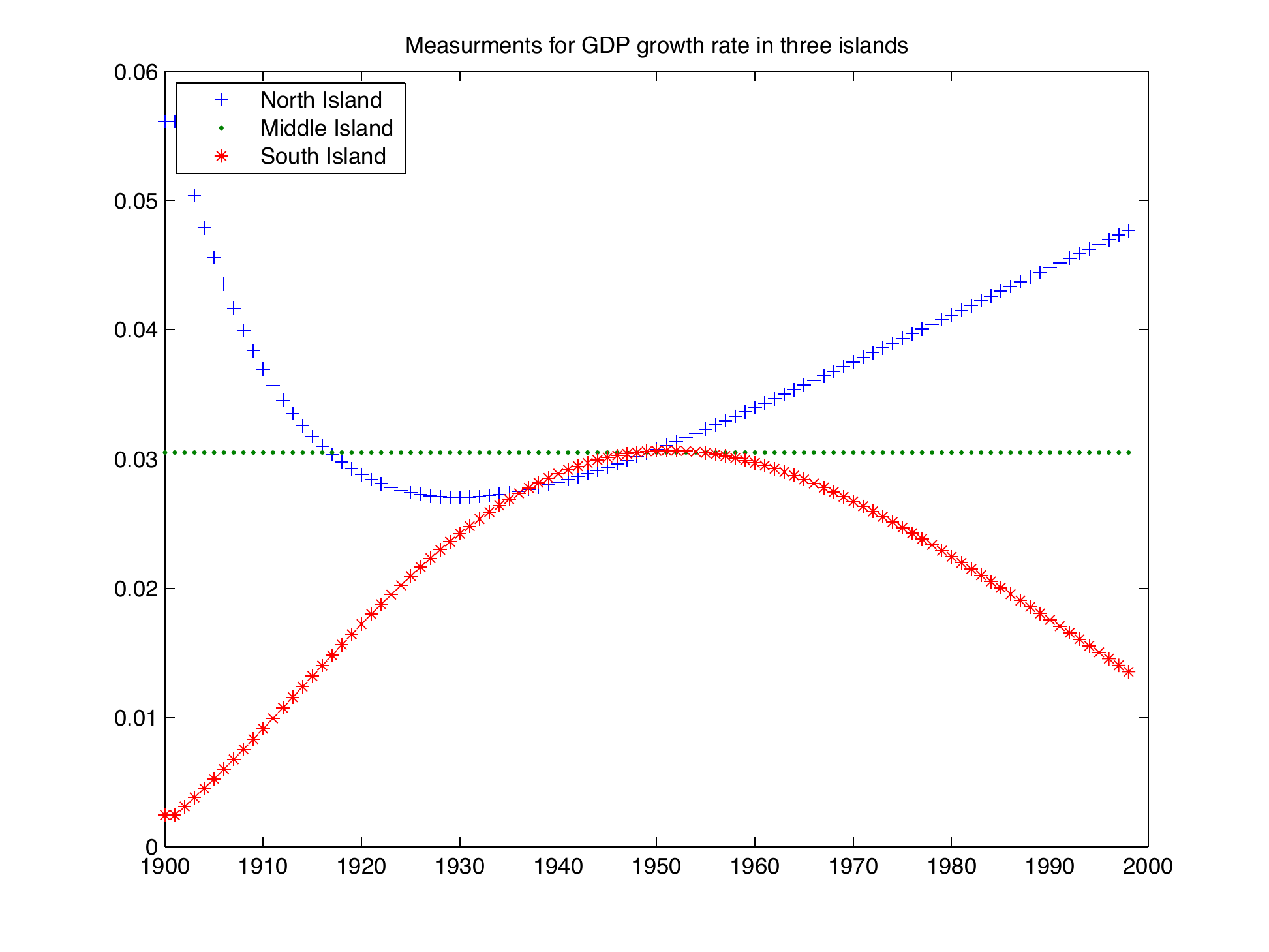}
                \caption{}
                \label{fig:2gdpgrowthrate}
        \end{subfigure}%
         %add desired spacing between images, e. g. ~, \quad, \qquad etc.
        \begin{subfigure}[b]{0.5\textwidth}
                \centering
                \includegraphics[width=\textwidth]{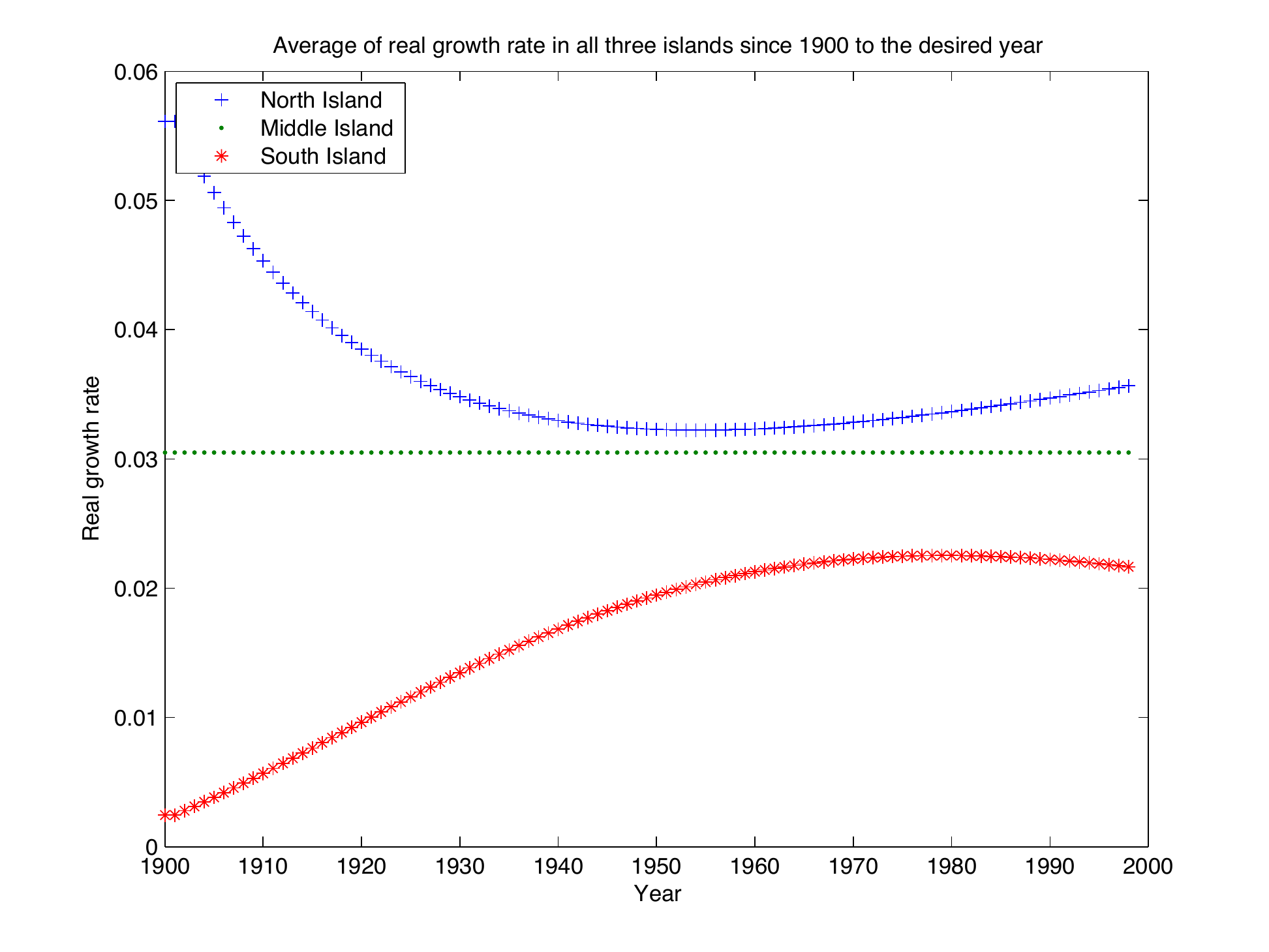}
                \caption{}
                \label{fig2gdpavereage}
        \end{subfigure}

        \caption[font=footnotesize]{This figure depicts real growth rates reported by central banks in each island. As we expect, since the pattern of improvements in productivity for different sectors is different in each island, the measurements for GDP suggest different growth rates (graph a). The surprising point is that when we add these growth rates for the whole period, and take average, we obtain different rates of growth for each country (graph b). The average growth rate for the whole period ending in 1998 is $3.6\%$ in North Island, $3.1\%$ in Middle Island, and $2.2\%$ in South Island.
This is of surprise since GDP is exactly the same in 1900 and 1998 for all three Islands.} \label{fig:2gdpgrowth}
\end{figure}

\begin{figure}
        \centering

        \begin{subfigure}[b]{0.5\textwidth}
                \centering
                \includegraphics[width=\textwidth]{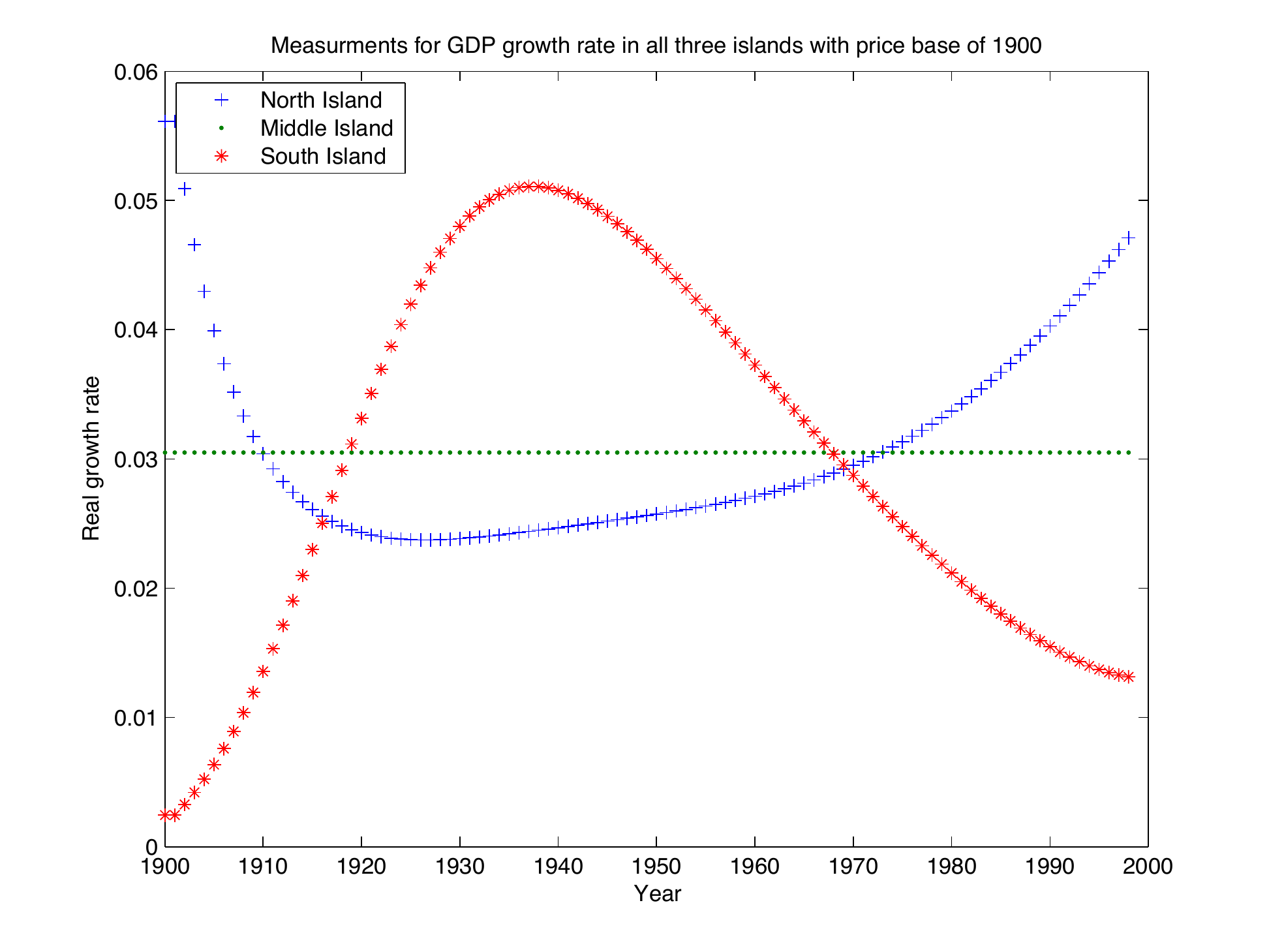}
                \caption{}
                \label{fig:2gdpgrowth1900}
        \end{subfigure}%
         %add desired spacing between images, e. g. ~, \quad, \qquad etc.
        \begin{subfigure}[b]{0.5\textwidth}
                \centering
                \includegraphics[width=\textwidth]{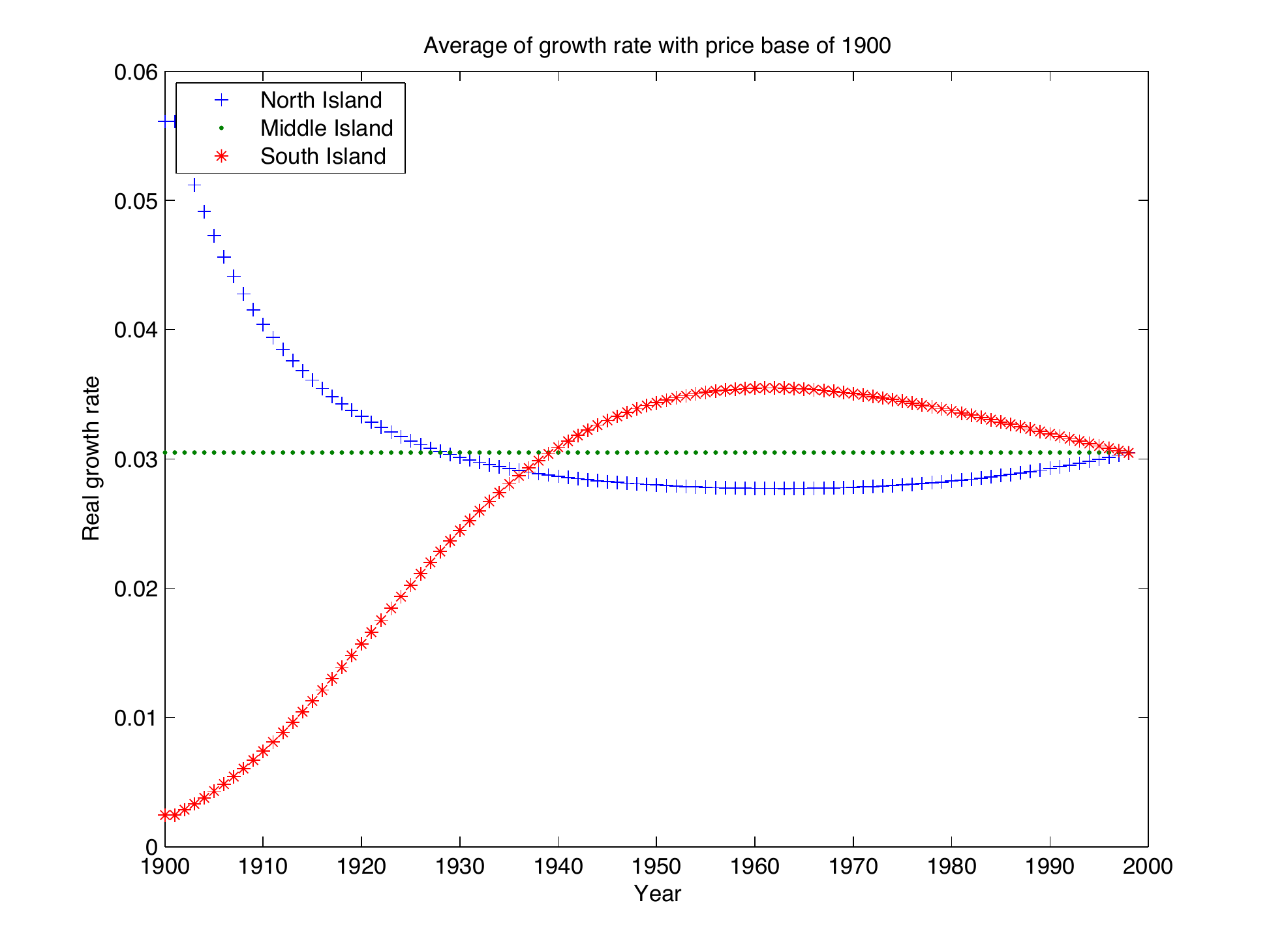}
                \caption{}
                \label{fig:2gdpaverage1900}
        \end{subfigure}

        \caption[font=footnotesize]{a) Real growth rates when for each year, 1900 is considered as the base year. b) Average of growth. As it can be seen as in the end of the period, production in all three islands merge together. The average growth reported by the central banks also merge.  So, if central banks keep 1900 as the base year, then we would not observe paradoxical results in this experiment. } \label{fig:2gdp1900}
\end{figure}

\begin{figure}
  \centering
    \includegraphics[width=0.5\textwidth]{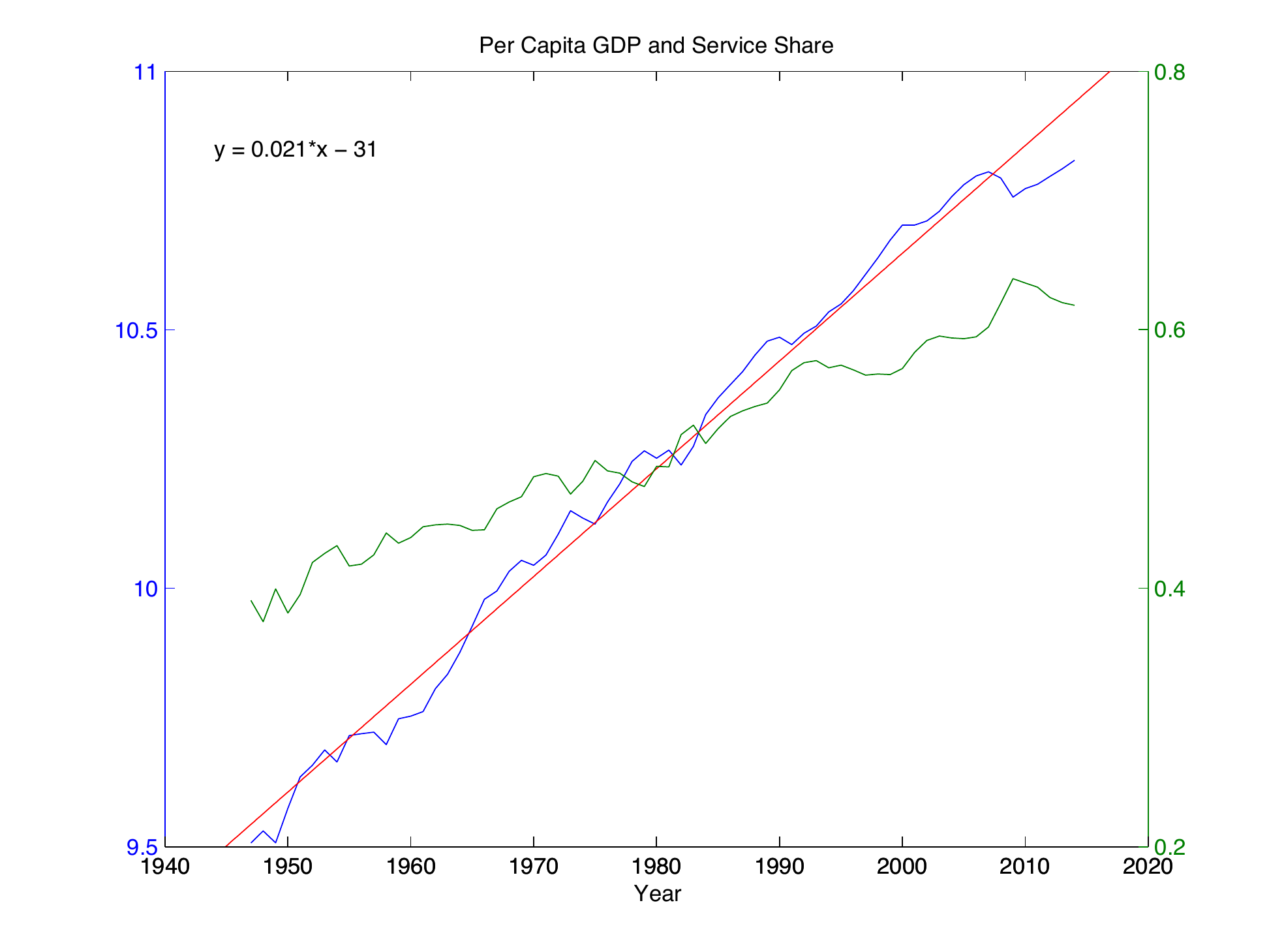}
  \caption{The green curve is service share of GDP in the US. The blue curve is per capita GDP chained 2009 US dollars. As it can be seen since the middle of 20th century share of services in GDP has grown from \%40 to \%60. Though growth of productivity in service side is far below the industrial side, increasing the share of service side has not resulted in decline of historical rate of growth for GDP. During this period per capita GDP has kept its historical growth rate of 2\%. It supports our claim that sustainable growth rate is a matter of measurement rather than being a real fact. Sources: GDP per capita from stlouisfed.org and service share from Bureau of Economic Analysis}\label{figgrowthess}
\end{figure}\label{figgdp}

\begin{figure}
  \centering
    \includegraphics[width=0.5\textwidth]{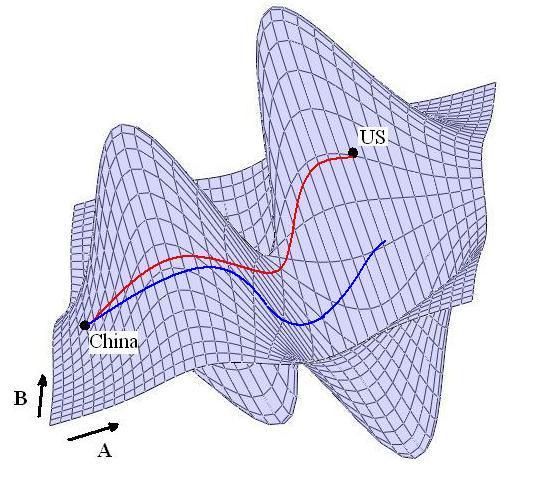}
  \caption{As we plan to find distance between economies of China and of the US we need to find  geodesic that pass through them and besides the local metric. We don't have an exact definition as geodesic. Countries however are bind to minimize utilities and thereby pass through special paths. China for sure uses a path that wont meet the US in the space of production (the blue path in our figure for example). We however can extrapolate the proper time that economy of China is as big as the US if we can measure the length of the path with local metric.}\label{figgeneralrelativity}
\end{figure}\label{figgeneralrelativity}

%--------------------------------------------------------------------

\begin{thebibliography}{}

%\cite{Baumol}
\bibitem{Kaldor}
   Kaldor, Nicholas 1957. "A Model of Economic Growth". The Economic Journal 67 (268): 591–624

%%%%%%%%%%%%%%%%%%%%%%%




%\cite{Baumol}
\bibitem{Crawford}
Crawford, I, and J. P. Neary 2008, “Testing for a Reference Consumer in International Comparisons of Living Standards”. American Economic Review, 98(4), 1731-1732.

%%%%%%%%%%%%%%%%%%%%%%%

%\cite{Baumol}
\bibitem{deaton2010}
  Deaton, A. and A. Heston 2010, “Understanding PPPs and PPP-based national accounts”. American Economic Journal: Macroeconomics 2010, 2:4, 1–35.
%%%%%%%%%%%%%%%%%%%%%%%


%\cite{Baumol}
\bibitem{deaton2010b}
 Deaton, A. 2010, “Price indexes, inequality, and the measurement of world poverty”. American Economic Review, 100:1, 5–34.
%%%%%%%%%%%%%%%%%%%%%%%


%\cite{Baumol}
\bibitem{deaton2014}
  Deaton, A., and Bettina A. 2014, “Trying to understand the PPPs in ICP2011: why are the results so different?”. NBER Working Paper no. 20244.
%%%%%%%%%%%%%%%%%%%%%%%
\bibitem{freenstra2013}
Feenstra, R. C.,  H. Ma, J. Peter Neary and D. Prasada Rao,  2013a "Who Shrunk China? Puzzles in the Measurement of Real GDP," The Economic Journal 123 (573), 1100-1129



%\cite{Baumol}
\bibitem{prasada}
  Prasada,R., A. Rambaldi and H. Doran 2010, “Extrapolation of Purchasing Power Parities using multiple benchmarks and auxiliary information: a new approach”. Review of Income and Wealth, 56, Special Issue 1, S59-S98.
%%%%%%%%%%%%%%%%%%%%%%%

%\cite{Baumol}
\bibitem{freenstra2009}
 Feenstra, Robert C., Hong Ma, and D.S. Prasada Rao 2009, “Consistent comparisons of real incomes across space and time”. Macroeconomic Dynamics, 13(S2), 169-193.
%%%%%%%%%%%%%%%%%%%%%%%

%\cite{Baumol}
\bibitem{feenstra2013}
 Feenstra, Robert C., Robert Inklaar, and Marcel P. Timmer 2013b, “The next generation of the Penn World Table”, NBER Working Paper 19255
%%%%%%%%%%%%%%%%%%%%%%%

%\cite{Baumol}
\bibitem{oulton2014}
 Oulton, N. 2015, “Understanding the space–time (in) consistency of the national accounts”. Economics Letters, 132, 21-23
%%%%%%%%%%%%%%%%%%%%%%%





%\cite{Baumol}
\bibitem{Baumol1966}
  Baumol,W. and W.~Bowen 1966
   "Performing arts: The economic dilemma.'' New York:Twentieth Century Fund.

%%%%%%%%%%%%%%%%%%%%%%%



%\cite{Baumol}
\bibitem{Baumol1989}
  Baumol, W., S. Batey Blackman, and E. Wolff. 1989
   "Productivity and American leadership: The long view.'' Cambridge, Mass.: MIT Press
%%%%%%%%%%%%%%%%%%%%%%%



%\cite{Baumol}
\bibitem{Baumol2012}
  Baumol, W. 2012
   "The Cost Disease
Why Computers Get Cheaper and Health Care Doesn’t.'' New Haven \& London, Yale University Press.

%%%%%%%%%%%%%%%%%%%%%%%
\bibitem{nationalcenter}
National Center for Education Statistics. 2010. Total tuition, room and board rates charged for full-time students in degree-granting institutions, by type and control of institution: Selected years, 1980–81 to 2008–09. In Digest of education statistics, 2009. Washington, D.C.: U.S. Department of Education. http://nces.ed.gov /fastfacts/display.asp?id=76.




%\cite{Baumol}
\bibitem{national}
 National Center for Public Policy and Higher Education. 2008. "Measuring up 2008: The nationa report card on higher education". San Jose, Calif, p.8, fig. 5. http://measuringup2008.highereducation.org/print/NCP- PHEMUNationalRpt.pdf.

%%%%%%%%%%%%%%%%%%%%%%%

%\cite{Baumol}
\bibitem{usbureau}
   U.S. Bureau of Labor Statistics 2009a. Legal services consumer price index. All Urban Consumers (Current Series) Database. http://data.bls.gov/PDQ /outside.jsp?survey=cu.
%%%%%%%%%%%%%%%%%%%%%%%

%\cite{Baumol}
\bibitem{usbureaub}
U.S. Bureau of Labor Statistics 2009b. Funeral expenses consumer price index. All Urban Consumers (Current Series) Database. http://data.bls.gov/PDQ /outside.jsp?survey=cu
%%%%%%%%%%%%%%%%%%%%%%%


%\cite{Baumol}
\bibitem{lu}
Lu, C., M. Schneider, P. Gubbins, et al. 2010. "Public financing of health in developing countries: A cross-national systematic analysis." Lancet 375:1375–1387
%%%%%%%%%%%%%%%%%%%%%%%



%\cite{Baumol}
\bibitem{fisher}
Fisher, I. 1927 [1967], The Making of Index Numbers: A Study of Their Varieties, Tests, and Reliability (3rd edition). Augustus M. Kelley, New York.
%%%%%%%%%%%%%%%%%%%%%%%




%\cite{Baumol}
\bibitem{balk}
Balk, B.M. 1995, Axiomatic price index theory: a survey. International Statistical Review,
63, 1, 69-93.
%%%%%%%%%%%%%%%%%%%%%%%

%\cite{Baumol}
\bibitem{Diewert}
Diewert, W. E., 1976 “Exact and Superlative Index Numbers.” Journal of Econometrics, 4: 115-46.
%%%%%%%%%%%%%%%%%%%%%%%



%\cite{Baumol}
\bibitem{feenstra2000}
Feenstra, R. C. and M. B. Reinsdorf (2000). “An Exact Price Index for the Almost Ideal Demand System.” Economics Letters, 66: 159-162.
%%%%%%%%%%%%%%%%%%%%%%%


%\cite{Baumol}
\bibitem{oulton2008}
Oulton, N., 2008, "Chain indices of the cost of living and the path-dependence problem: An epirical solution", Journal of Econometrics,
%%%%%%%%%%%%%%%%%%%%%%%




%\cite{Baumol}
\bibitem{Ngai}
Ngai, L. R. and Pissarides, C. 2007 “Structural Change in a Multisector Model of Growth.” American Economic Review, 97(1), 429-443.
%%%%%%%%%%%%%%%%%%%%%%%



%\cite{Baumol}
\bibitem{Foellmi}
Foellmi, R. and Zweimueller, J. 2008, “Structural Change, Engel’s Con- sumption Cycles and Kaldor’s Facts of Economic Growth”, Journal of Monetary Economics, 55(7), 1317-1328
%%%%%%%%%%%%%%%%%%%%%%%




%\cite{Baumol}
\bibitem{Herrendorf}
Herrendorf, B., R. Rogerson, and A. Valentinyi 2013, "Growth and structural transformation", In P. Aghion and S. Durlauf (Eds.), Handbook of Economic Growth, Volume 2B, Chapter 6, pp. 855–941. North-Holland.
%%%%%%%%%%%%%%%%%%%%%%%


%\cite{Baumol}
\bibitem{Boppart}
Boppart, T. 2014, “Structural Change and the Kaldor Facts in a Growth Model With Relative Price Effects and Non?Gorman Preferences,” Economet- rica, 82, 2167–2196.

%%%%%%%%%%%%%%%%%%%%%%%





%\cite{Baumol}
\bibitem{Hosseiny2016}
Hosseiny, A., Gallegati, Mauro (2016), to be appeared, "The Role of Intensive and Extensive Variables in a Soup of Firms in Economy to Address Long Run Prices and Aggregate Data"
%%%%%%%%%%%%%%%%%%%%%%%





%\cite{Baumol}
\bibitem{cohen}
 Cohen A.J, G.C. Harcourt 2003 "Retrospectives: Whatever happened to the Cambridge capital theory controversies?", Journal of Economic Perspectives 17:1, 199-214





%%%%%%%%%%%%%%%%%%%%%%%
\captionsetup[figure]{labelfont=bf,textfont={it}}
\end{thebibliography}
\end{document}